\documentclass[twocolumn]{aastex62}

\newcommand{\Nstars}{353 }
\newcommand{\Nstarstot}{16762 }
\graphicspath{{./}{figures/}}
\hypersetup{linkcolor=red,citecolor=green,filecolor=cyan,urlcolor=magenta}
\received{July 15, 2019}
\revised{July 25, 2019}
\accepted{July 27, 2019}
\submitjournal{ApJS}
\shorttitle{bRing Variables}
\shortauthors{Mellon et al.}
\begin{document}

\title{Bright Southern Variable Stars in the bRing Survey}

\correspondingauthor{Samuel N. Mellon}
\email{smellon@ur.rochester.edu}

\author[0000-0003-3405-2864]{Samuel N. Mellon}
\affil{Department of Physics \& Astronomy, University of Rochester, 500 Wilson Blvd., Rochester, NY 14627, USA}

\author[0000-0003-2008-1488]{Eric E. Mamajek}
\affil{Jet Propulsion Laboratory, California Institute of Technology, M/S 321-100, 4800 Oak Grove Drive, Pasadena, CA 91109, USA}
\affil{Department of Physics \& Astronomy, University of Rochester, 500 Wilson Blvd., Rochester, NY 14627, USA}

\author[0000-0001-7797-3749]{Remko Stuik}
\affil{Leiden Observatory, Leiden University, PO Box 9513, 2300 RA Leiden, The Netherlands}

\author[0000-0001-9229-8315]{Konstanze Zwintz}
\affil{Institut f{\"u}r Astro- und Teilchenphysik, Universit{\"a}t Innsbruck, Technikerstrasse 25/8, 6020 Innsbruck, Austria}

\author[0000-0002-7064-8270]{Matthew A. Kenworthy}
\affil{Leiden Observatory, Leiden University, PO Box 9513, 2300 RA Leiden, The Netherlands}

\author[0000-0003-4787-2335]{Geert Jan J. Talens}
\affil{Institut de Recherche sur les Exoplan\`etes, D\'epartement de Physique, Universit\'e de Montr\'eal, Montr\'eal, QC H3C 3J7, Canada}

\author[0000-0002-2487-4533]{Olivier Burggraaff}
\affil{Leiden Observatory, Leiden University, PO Box 9513, 2300 RA Leiden, The Netherlands}
\affil{Institute of Environmental Sciences (CML), Leiden University, PO Box 9518, 2300 RA Leiden, The Netherlands}

\author[0000-0002-4272-263X]{John I. Bailey, III}
\affil{Department of Physics, University of California at Santa Barbara, Santa Barbara, CA 93106, USA}

\author[0000-0003-3812-2436]{Patrick Dorval}
\affil{Leiden Observatory, Leiden University, PO Box 9513, 2300 RA Leiden, The Netherlands}

\author[0000-0002-1520-7851]{Blaine B. D. Lomberg}
\affil{South African Astronomical Observatory, Observatory Rd, Observatory Cape Town, 7700 Cape Town, South Africa}
\affil{Department of Astronomy, University of Cape Town, Rondebosch, 7700 Cape Town, South Africa}

\author[0000-0002-4236-9020]{Rudi B. Kuhn}
\affil{South African Astronomical Observatory, Observatory Rd, Observatory Cape Town, 7700 Cape Town, South Africa}

\author[0000-0002-6194-043X]{Michael J. Ireland}
\affil{Research School of Astronomy and Astrophysics, Australian National University, Canberra, ACT 2611, Australia}

\begin{abstract}
Besides monitoring the bright star $\beta$ Pic during the near transit event for its giant exoplanet, the $\beta$ Pictoris b Ring (bRing) observatories at Siding Springs Observatory,  Australia and Sutherland, South Africa have monitored the brightnesses of bright stars ($V$ $\simeq$ 4--8 mag) centered on the south celestial pole ($\delta$ $\leq$ -30$^{\circ}$) for approximately two years. 
Here we present a comprehensive study of the bRing time series photometry for bright southern stars monitored between 2017 June and 2019 January. 
Of the \Nstarstot stars monitored by bRing, \Nstars of them were found to be variable. 
Of the variable stars, 80\% had previously known variability and 20\% were new variables.
Each of the new variables was classified, including 3 new eclipsing binaries (HD 77669, HD 142049, HD 155781), 26 $\delta$ Scutis, 4 slowly pulsating B stars, and others.
This survey also reclassified four stars based on their period of pulsation, light curve, spectral classification, and color-magnitude
information.
The survey data were searched for new examples of transiting circumsecondary disk systems, but no candidates were found.
\end{abstract}


\section{Introduction} \label{sec:intro} 

Over the past two decades, several wide-field, ground- and space-based surveys have contributed countless hours of observations in the night sky \citep[e.g., KELT, MASCARA, NASA's Kepler and K2 Space missions:][]{Pepper07,Talens17,Kepler_main,Kepler_K2_Main}. The primary goal of these surveys has been the discovery of exoplanets, with each having a number of significant successes \citep[e.g.,][]{Oberst17,TalensM1}. A secondary result from these surveys has been the discovery and characterization of variable stars \citep[e.g.,][]{Collins18,Burggraaff18}.

Variable stars form the corner stone of much of the knowledge about our universe such as asteroseismology \citep[e.g.,][]{Zwintz14,Zwintz14c}, stellar gyrochronology and rotation \citep[e.g.,][]{Hartman10,Gallet13,Cargile14,Mellon17}, classical Cepheids as standard candles for distance \citep[e.g., ][]{Groenewegen18}, and eclipsing systems \citep[e.g.,][]{Mellon17,Moe18,Collins18}. In addition to the exoplanet surveys, dedicated variable star observatories and online catalogues have fueled research in these areas \citep[e.g., ASAS, AAVSO, ASAS-SN, OGLE:][]{Pojmanski02,Watson06,Shappee14, Udalski08}. Physical properties of stellar systems can be constrained from the period and amplitude of the observed variability, such as the composite sinusoidal variability observed in the $\delta$ Scuti star $\beta$ Pictoris  \citep{Mekarnia17,Zwintz19}.

In 2017, the $\beta$ Pictoris b Ring (bRing) instruments (located in South Africa and Australia) were constructed and brought online to observe the 2017-2018 transit of the $\beta$ Pictoris b Hill sphere \citep{Stuik17, Mellon19AAS,Kalas19}. While observing $\beta$ Pictoris, bRing captured nearly continuous photometry of 10,000+ bright stars ($V$ $\sim$ 4--8 mag) in the southern sky ($\delta$ $\leq$ -30$^{\circ}$). In addition to the study of the $\beta$ Pictoris b Hill sphere, the bRing survey has contributed the discovery of $\delta$ Scuti pulsations in the A1V star HD 156623 \citep{Mellon19}, the study of $\beta$ Pictoris' $\delta$ Scuti pulsations \citep{Zwintz19}, and the discovery of the retrograde hot Jupiter MASCARA-4 b/bRing-1 b \citep{Dorval19}.

In this work, we took a similar approach to the MASCARA survey of the northern sky \citep{Burggraaff18} and searched for periodic variations in the bRing data. This survey was also sensitive to evidence of transiting circumplanetary systems like ``J1407'' \citep[V1400 Cen;][]{Mamajek12}, or other circumsecondary disks, however, none were found. Section 2 of this work describes the data from both the South African bRing (bRing-SA) and Australian bRing (bRing-AU) stations. Section 3 details the analysis used to identify and characterize both the regular and irregular variables in the data. Section 4 provides tables and discussion of each type of variable found in cross-correlation with the VSX catalogue \citep{Watson06} and others.

\section{Data}\label{sec:data}  

The data in this work were collected between 2017 June and 2019 January by the bRing-SA and bRing-AU stations. Each station had two stationary cameras; one camera faced southeast (Az = 150$^{\circ}$; SAE and AUE) and the other southwest (Az = 230$^{\circ}$; SAW and AUW). Each camera had an FLI 4008 $\times$ 2672 pixel CCD and f = 1.4mm Canon wide-angle lens, which resulted in a total field of view of 
74$^{\circ}$ $\times$ 53$^{\circ}$ with a pointing optimized for $\beta$ Pictoris ($\delta$ $\simeq$ -53$^{\circ}$). Exposure times were alternated between 6.4s and 2.54s; these were subsequently co-added and binned to 5 minute samplings and saved to disk \citep{Stuik17,Talens18}.

Due to bRing's large pixel size ($\sim$1 arcmin$^2$), blending was a significant issue for bRing. Blending was evaluated by comparing the relative brightnesses of stars located within the same bRing inner aperture (radius = 2.5') as the target star \citep[nearby stars evaluated with the ASCC catalogue;][]{Kharchenko01}. For stars with previously known variability, blending was ignored if the original period was detected, but considered if a second period was detected or dominated the expected period. If a second period dominated the expected period in a blended star, the star was reanalyzed at the original expected period. If a star showed signs of variability in our light curves and had been previously unidentified as a variable in other surveys, blending was required to be 0 to be considered a detection. Ultimately, \Nstarstot stars were analyzed for this work. The stars listed in this work as new variables had no evidence of blending in their light curves.

On average, a star had observations spanning over 300 days (each star ideally received 21 hours of continuous coverage per day); the average star had $\sim$ 20,000 five minute binned data points over the entire observing window combined from all 4 cameras. More information on the bRing observing strategy and data calibration can be found in \citet{Stuik17, Talens18}. In conjunction with this work, the camera .FITS files for each star \citep[as described in][]{Stuik17} were published in a Zenodo repository \citep{MellonZenodo19}.

In interpreting the nature of the variability, $BV$ photometry was drawn from the ASCC-2.5 catalog \citep[][]{Kharchenko01} and spectral types were drawn from the literature, with most types
taken from the Michigan Spectral Survey of classifications from objective-prism plates \citep{Houk75,Houk78,Houk82}. \citet{Houk97} has shown that for $V$ $<$ 8 mag stars classified as dwarf luminosity class in the Michigan Spectral Survey, for a given 2D spectral type the intrinsic color spread rms in $B-V$ is $\sim$0.03-0.04 mag and the intrinsic spread in absolute $V$ magnitudes is $\sim$0.4-0.5 mag, with distributions suggesting negligible contamination by more evolved giants and supergiants. The Michigan classifications for the variable stars have quality flags of
1 (61.6\%), 2 (28.9\%), 3 (7.0\%), 4 (0.6\%), with the 93\%\, flagged
as quality 1 and 2 considered the "higher-quality" classifications \citep{Houk78,Houk97}.

\section{Analysis}\label{sec:analysis}  

The 5 minute binned data points from bRing were automatically calibrated and detrended for temporal and spatial effects from the observations \citep[e.g., clouds, intra-pixel variations;][]{Stuik17,Talens18}. Using an internal custom pipeline detailed in \citet{Mellon19}, these data were downloaded from the bRing server and further detrended for sidereal and lunar systematics as well as astrometric and color systematics. This routine also includes a Barycentric correction. In addition to the detrending from previous works, we attempted to preserve the ansatz period prior to detrending by including an additional step adopted from \citet{Burggraaff18}. The data for each star from each of the 4 bRing cameras were treated individually and then median-combined after detrending.

\subsection{Identifying the Ansatz Period} \label{ansatz} 

The time series photometry data was analyzed using the reduction pipeline previously used and described in \citet{Mellon19},
with a modification based on the study by \citet{Burggraaff18}. The step adopted from \citet{Burggraaff18} to improve upon the process from \citet{Mellon19} was the initial identification and removal of an ansatz period from the data prior to detrending. The goal of this step was to preserve any real and significant periods from being affected by the detrending process. To find the ansatz period, a normalized Lomb-Scargle periodogram \citep{Scargle82,Press92} was generated using the \textsc{astropy} \citep{astropy} library. Next, a Python routine was written using tools available in the \textsc{scipy} \citep{Scipy}, \textsc{numpy} \citep{Numpy}, and \textsc{astropy} packages to identify the strongest periods in the periodogram. These periods were then compared to the well-studied sidereal and lunar systematics present in the bRing data \citep[the origins of these systematics and methods for removing them are thoroughly discussed in][]{Stuik17,Talens18,Burggraaff18,Mellon19}. The strongest period that was not within 5\% of one of these systematics (or the corresponding harmonics and aliases to order 5) was accepted as the ansatz period, fit with a sine, and removed from the lightcurve. This information was stored and was added back in after detrending.

\subsection{Detrending and Measurement of Variable Star Parameters} 

\begin{figure}
\begin{center}
\plotone{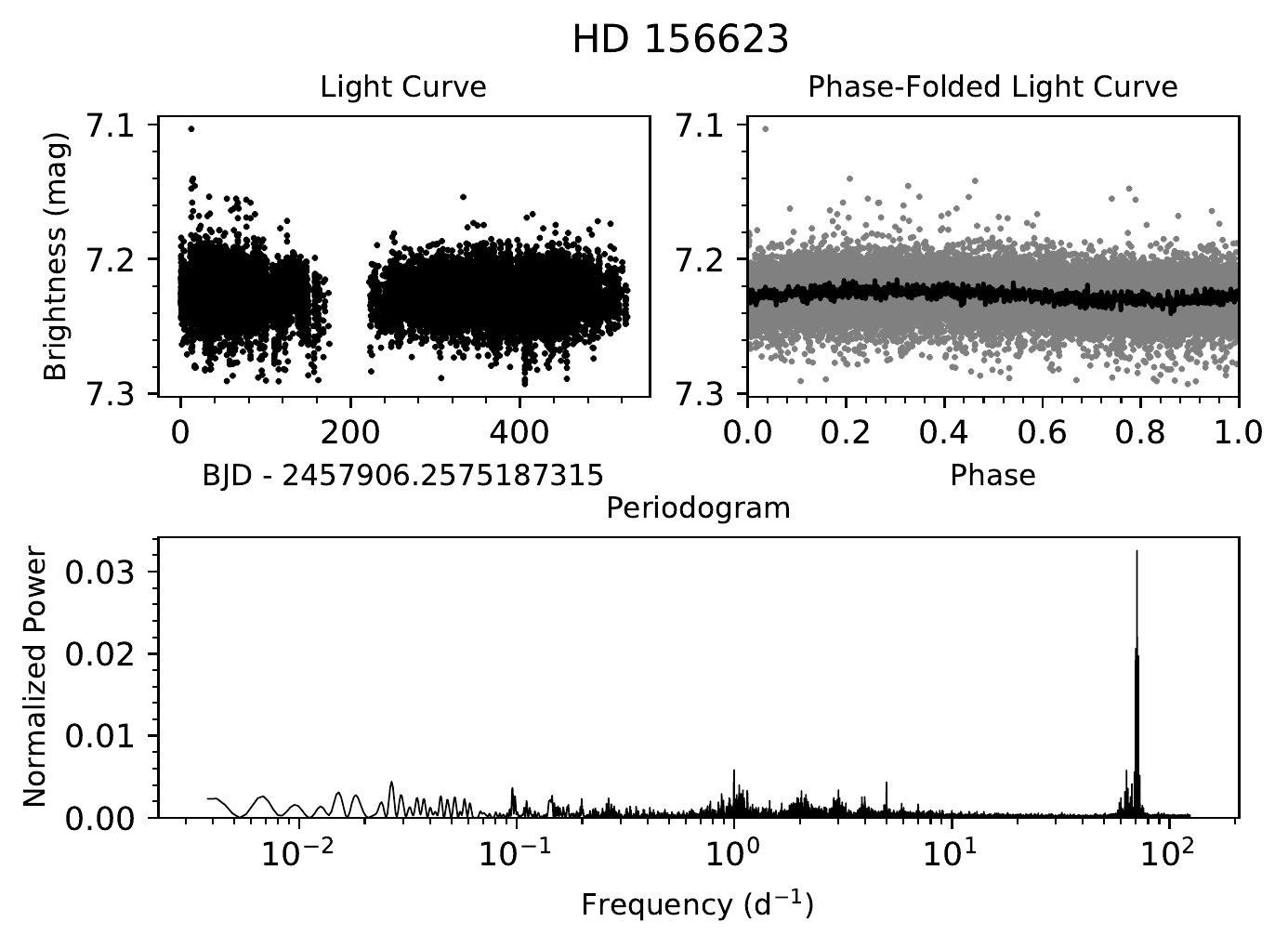}
\caption{An example plot of the $\delta$ Scuti HD 156623. The top left panel contains the light curve of the star. The top right panel contains the light curve of the star (gray dots) phase-folded on the primary period with a running median fit (solid curve). The bottom panel contains the normalized LS periodogram.}
\label{plotex}
\end{center}
\end{figure}

The detrending routine used after the removal of ansatz period is described in \citet{Mellon19} and is summarized in this work. First, an astrometric correction was applied to remove data points that deviated $>$ 3$\sigma$ from the mean path of the star on the CCD. Then, the time-series was adjusted to the Barycentric reference frame and a second-order CCD color correction was applied to the data. The best ansatz signal was then determined and temporarily removed from the data. Next, a median-binning routine was used to significantly reduce the strength of the lunar and sidereal systematic signals. After detrending, the ansatz signal was added back into the light curve and a composite light curve was generated from the 4 camera light curves using a median-alignment. A new periodogram was calculated from this composite light curve. Finally, a plot of the composite light curve, a periodogram, and a phase-folded light curve on the most-likely variability period was generated for analysis. These plots were used to identify variables in the data by eye. An example is plotted in Figure \ref{plotex} for the $\delta$ Scuti HD 156623. The plots generated for this work were included in the same Zenodo repository as the data \citep{MellonZenodo19}.

The measurements and information used to construct the tables of variable stars (see \S \ref{discussion}) were also generated. The best periods for variables were taken directly from the composite light curve periodograms and verified by comparing measurements from the 4 independent camera data sets. A previously unseen frequency was accepted if it was detected in at least one camera from each site. Uncertainties for the frequencies were measured from the standard deviation in the detected frequencies; amplitude uncertainties were calculated using methods from \citet{Montgomery99}. We compared the \citet{Montgomery99} frequency uncertainty measurements to the measurements from using the 4 camera data sets and found the uncertainties were typically underestimated by a factor of $\sim$5. This is expected as the uncertainties from \citet{Montgomery99} were noted in their work as lower limits on the errors in these measurements.

\begin{figure*}
\centering
\plotone{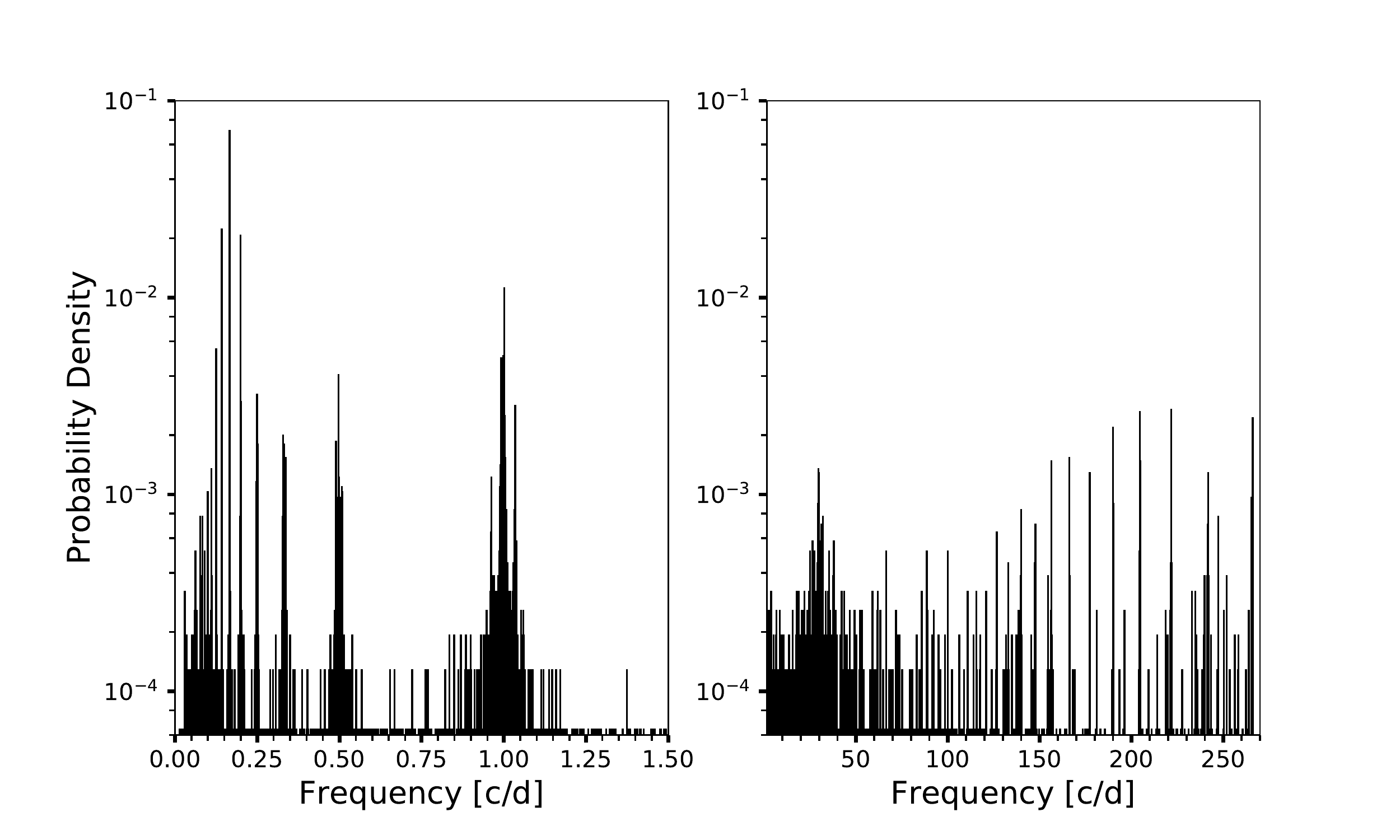}
\caption{The probability density plot of the strongest frequencies in the periodograms of all the stars in this study. The left panel focuses on low frequency (0 -- 1.5 d$^{-1}$) pulsations with clear contributions at 1 d$^{-1}$ and its aliases. The right panel contains higher frequencies ($>$ 1.5 d$^{-1}$) and has notable contributions at $\sim$ 30 d$^{-1}$, 100 d$^{-1}$, 200 d$^{-1}$ and their beats and aliases. The noise floor of the plot is at a probability density of $\sim$10$^{-4}$. It is worth noting the low frequency systematics are at least an order of magnitude stronger than the high frequency systematics, which are themselves an order of magnitude stronger than the noise floor.}
\label{hist}
\end{figure*}

\subsection{False Positives\label{FP}} 

The strongest periodogram frequencies from the stars in this study were used to identify remaining low frequency ($f$ $<$ 1.5 d$^{-1}$) and high frequency ($f$ $>$ 1.5 d$^{-1}$) false positives due to systematics in the bRing system. To do this, a density plot of the strongest frequencies was generated with bin sizes of 0.01 d$^{-1}$ (Figure \ref{hist}). The left panel focuses on the low frequency false positives, which have been discussed thoroughly in \citet{Stuik17,Talens18,Burggraaff18,Mellon19}. The high frequency systematics were observed to be more numerous and scattered, but are weaker by a factor of $\sim$10 compared to the low frequency systematics and are roughly a factor of 10 above the noise floor ($\simeq$ 10$^{-4}$) of the plot. Possible sources include the 288 d$^{-1}$ (5 minute) sampling frequency of bRing and its beats/aliases and electromagnetic interference within bRing. The bRing detrending routines are continuing to be internally developed to minimize the effects of these dominating systematics.

The low frequency systematics posed the largest problem due to the majority of the variables in this survey having real frequencies in this regime. They are clearly dominated by the sidereal cycle and its aliases; the large peaks that pick up around 0.167 $d$ is due to the ansatz routine not picking up frequencies at that harmonic. We were careful when reporting frequencies as real when they were within 0.1 d$^{-1}$ of these frequencies. For example, if independent evidence of variability existed for these frequencies near a systematic \citep[e.g., the eclipsing binary V397 Pup with a 3.00402 day period;][]{Watson06}, they were accepted as real. However, potential new variables could have been missed due to the lack of a sophisticated means of independent verification or imperfections in the detrending or ansatz routines. The high frequency systematics were only applicable to the $\delta$ Scuti candidates due to their high frequency regime, however, the systematics were not an issue for the $\delta$ Scuti primary frequencies detected in this study.

\subsection{Performance Analysis} 

\begin{figure*}
\centering
\plotone{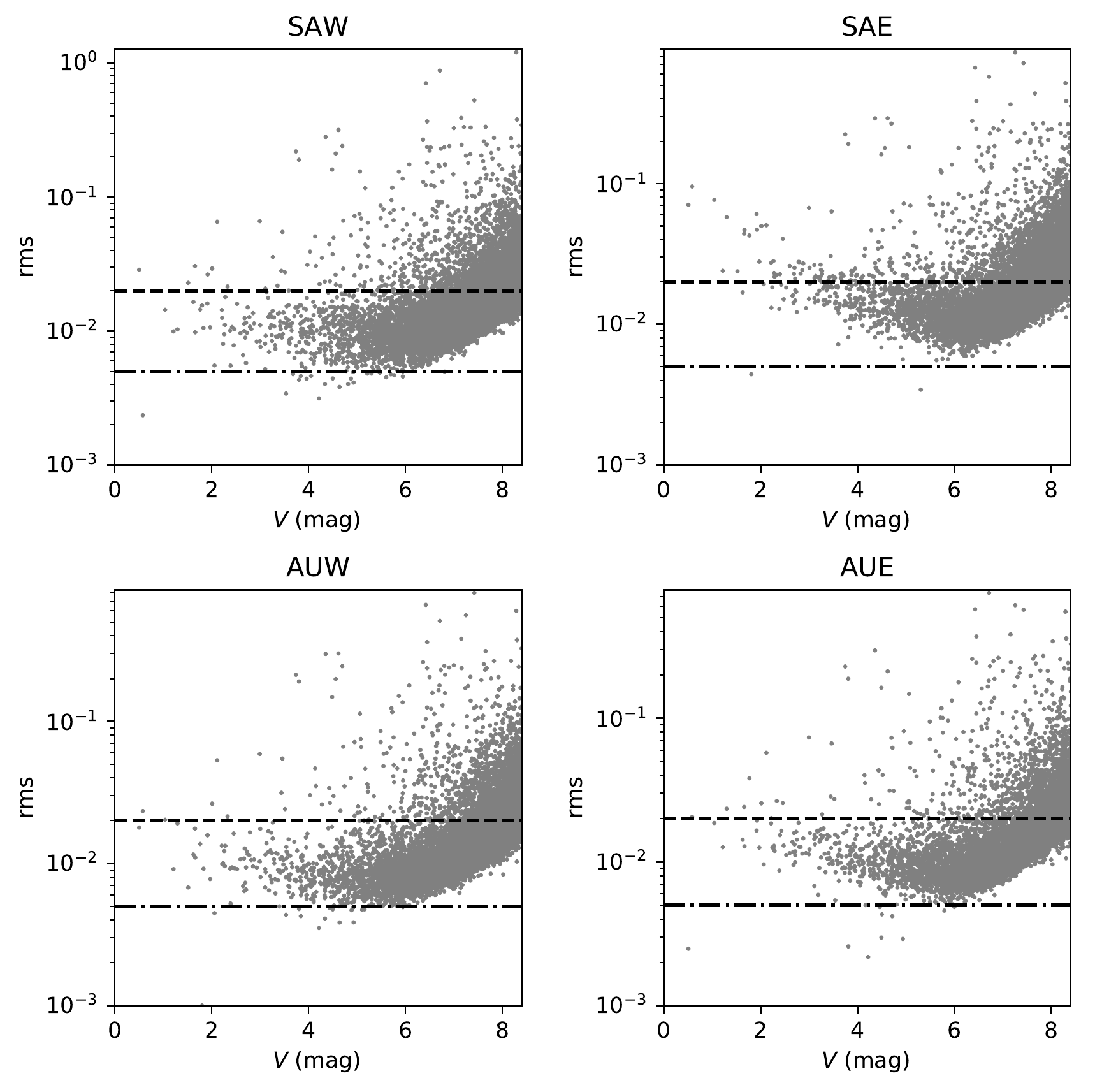}
\caption{Plots showing the rms (gray dots) in each of the cameras for the 16762 stars in this survey. The horizontal dashed line represents 2\% scatter. The horizontal dot-dash line represents the combined noise floor estimated by visual inspection to be at an rms of around 0.005.}
\label{rms}
\end{figure*}

The sample from this study was also used to study the performance of bRing. In Figure \ref{rms}, the rms for each post-detrending star was plotted in gray against the catalogue magnitude of the star. For each camera, $\sim$14\% of the stars performed better than 1\% and $\sim$70\% stars performed better than 2\% (dashed line). The results here are similar to the results from \citet{Talens17,Talens18}. 

By visual inspection, it is clear that a combined noise floor (plotted as a horizontal dot-dash line in Figure \ref{rms}) exists in all 4 cameras at an rms of about 0.005. The region brighter than $V$ $\simeq$ 5.5 mag is dominated by this combined noise floor term. Contributing terms to this noise floor include scintillation noise \citep[estimated to be around 10$^{-4}$ at both sites via Young's approximation:][]{Young67,Osborn15}, noise contributed from the calibration and detrending, as well as other noise sources such as read noise and dark current. This noise floor level matches the expected photometric precision for bright stars in bRing, indicating that the detrending routine used in this work was successful \citep{Stuik17,Talens17,Talens18}. The fainter region was dominated by the shot noise and sky noise contributions. Overall, bRing performed as expected at the bright end and performed well for stars at the faint end, which made a complete survey of all the stars in the bRing data possible despite lingering systematics.

\section{Results \& Discussion \label{discussion}} 

We detected \Nstars variable stars in the bRing survey.
We used the VSX\footnote{The VSX catalog is regularly updated at \url{https://www.aavso.org/vsx/}.} catalogue, Vizier\footnote{\url{https://vizier.u-strasbg.fr/}}, and SIMBAD\footnote{\url{http://simbad.u-strasbg.fr/simbad/}} web services to identify previously known or candidate variables \citep{Watson06, Ochsenbein00,Wenger00}.
The periods reported for previously known variables were then compared to the periods detected with bRing.
Stars that had no mention as variable stars in these databases, or suspected variables that did not have quoted 
periods in any source, are reported here as new periodic variables. 
Of the 284 previously known variables in this survey, the bRing periods were found to be consistent for 62\% of the stars.
The majority of the inconsistent periods were $\delta$ Scutis or long period variables.
bRing could simply be detecting a more significant period or alias for the $\delta$ Scutis due to their multi-periodic nature that requires further study to disentangle (out of the scope of this work).
The long period variables typically had low precision measurements of the period, leading to the inconsistencies observed between previous measurements and bRing measurements.

bRing detected 71 variables that had not been previously flagged as known or candidate variables (including the 17 irregular variables observed by bRing).
bRing was also able to reclassify four stars based on their newly measured period, light curve shape, and spectral classification.
The remainder of the stars showed no detectable or independent signs of variability down to the $\sim$1 mmag level. 
These stars are tabulated by variable classification in the following subsections.

\begin{figure*}
\centering
\plotone{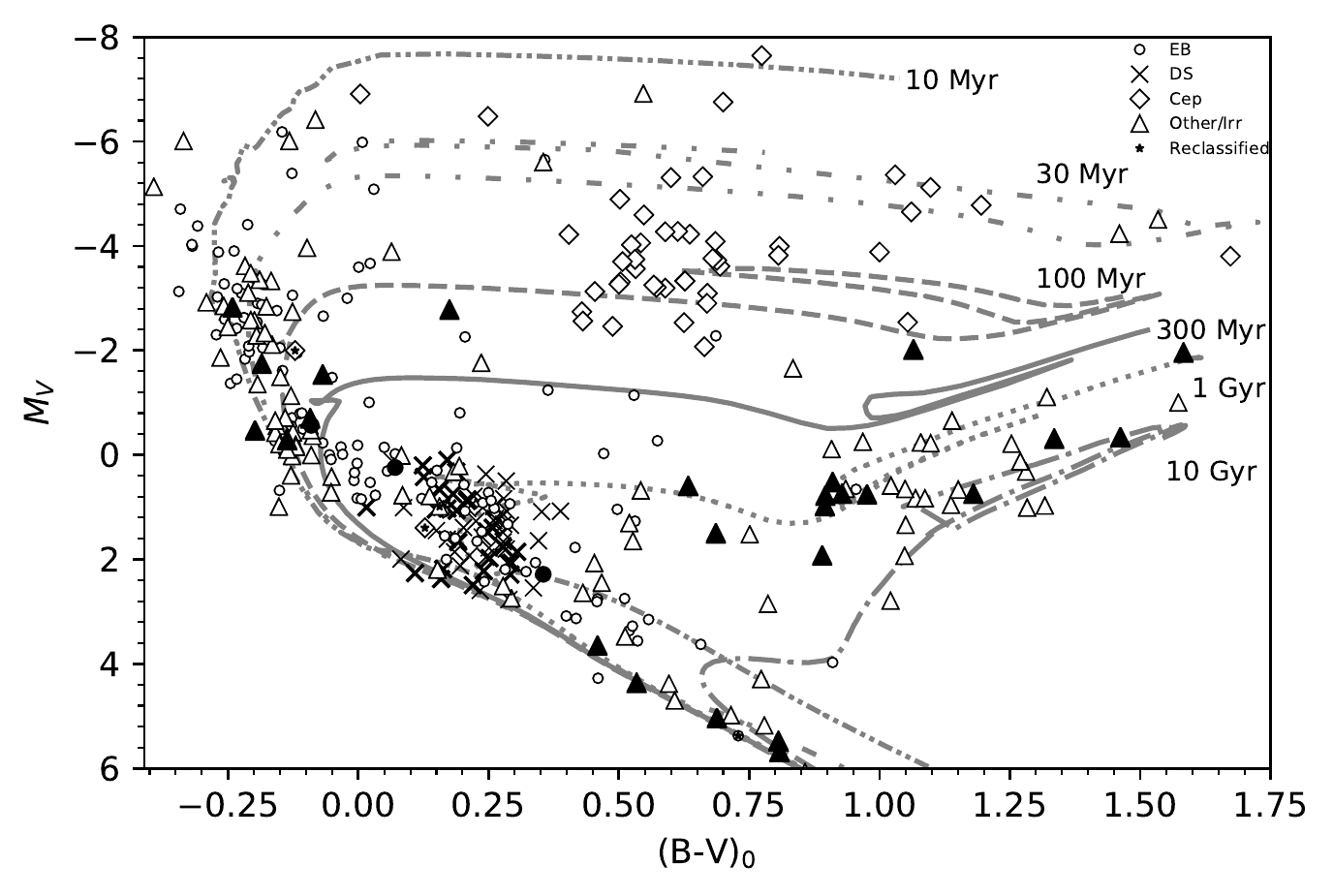}
\caption{A color-magnitude diagram of the variables in this work. Types of variables are symbol-coded with respect to the tables presented in \S \ref{discussion}. The symbols for previously identified variables are outlined in black and newly identified variables are solid black. The 4 reclassified variables in this work are denoted with a black star. Several solar composition PARSEC isochrones are overlayed \citep{Bressan12,Marigo17}. Two of the stars had reddening values that placed them outside the scope of this plot; these stars were not plotted in order to better focus on the majority of the stars HD 85871 ($(B-V)_0$ = 2.32, $M_V$ = -3.45) and HD 30551 ($(B-V)_0$ = 2.63, $M_V$ = -3.99). Another star (HD 69342) was not included since a distance could not be determined for this star.}
\label{cmd}
\end{figure*}

The color-absolute magnitude positions of the \Nstars variable stars are plotted in Figure \ref{cmd}. The different types of variables are symbol coded with respect to the tables they inhabit in \S \ref{discussion}. Previously known variables are outlined in black while newly identified variables are solid black. For the 4 stars reclassified in this work, a small black star was placed on top of their respective symbol. The SIMBAD service was queried for Johnson $BV$ photometry \citep{ESA97,Kharchenko01}, Galactic coordinates $l$ and $b$, and in the vast majority of cases, either a Gaia DR2 or Hipparcos parallaxes \citep[][]{Ochsenbein00,GaiaDR2,vanLeeuwen07}. We queried the most recent 3D reddening maps from the STILISM\footnote{\url{https://stilism.obspm.fr/}} program to de-redden the $(B-V)$ colors \citep{Capitanio17, Lallement18}. Following \citet{Mellon19}, we adopted the ratio of total to selective extinction to be $A_V$/E($B-V$) $\simeq$ 3.07 + 0.167\,$(B-V)_o$, which is
an adequate approximation over the intrinsic color interval -0.32 $<$ $(B-V)_o$ $<$ 1.5. 
Solar composition PARSEC isochrones \citep{Bressan12,Marigo17} were overlayed for several ages; these were generated using the CMD 3.3 Input tool\footnote{\url{http://stev.oapd.inaf.it/cgi-bin/cmd}}. The color-absolute magnitude
parameters calculated for Figure \ref{cmd} are tabulated in Table \ref{cmdtab} in the Appendix.

\subsection{Cepheid Variables}\label{sec:ceph} 
\startlongtable
\begin{deluxetable*}{c|c|c|c|c|c|c|c|c|c|c}
\tablecaption{Previously Classified Cepheid Variables Detected With bRing \label{cepht}}
\tablehead{\colhead{Name} &\colhead{HD} & \colhead{P} & \colhead{$\sigma_{P}$} & \colhead{A} & \colhead{$\sigma_{A}$} &\colhead{VSX ID} & \colhead{P$_{VSX}$} & \colhead{$V$} & \colhead{SpT} & \colhead{Ref}\\
\colhead{...} &\colhead{...} & \colhead{d} & \colhead{d} & \colhead{mmag}& \colhead{mmag}& \colhead{...} & \colhead{d} & \colhead{mag} & \colhead{...}& \colhead{...}}
\startdata
bet Dor&37350&9.844&0.007&157.8&1.6&13671&9.843&3.80&F6Ia&1\\
AP Pup&65592&5.085&0.003&127.8&2.6&26671&5.084&7.34&F8II&2\\
AX Vel&68556&2.592&0.001&82.8&1.5&37493&2.593&8.14&F6II&3\\
%
%
AH Vel&68808&4.229&0.033&87.5&1.2&37478&4.227&5.73&F7IB/II&3\\
RS Pup&68860&41.193&0.429&242.5&5.9&26613&41.443&7.00&F8Iab&2\\
V Car&72275&6.699&0.003&143.0&1.7&5758&6.697&7.30&F8Ib/II&1\\
RZ Vel&73502&20.482&0.067&264.4&4.0&37434&20.398&7.15&G1Ib&3\\
SW Vel&74712&23.381&0.067&251.7&3.9&37439&23.407&8.30&F8/G0Ib&3\\
SX Vel&74884&9.537&0.013&169.7&2.6&37440&9.550&8.34&F8II&3\\
BG Vel&78801&6.928&0.01&97.0&1.4&37501&6.924&7.68&F7/F8II&3\\
V Vel&81222&4.366&0.004&144.2&1.8&37421&4.371&7.57&F8II&1\\
I Car&84810&35.688&0.060&167.4&1.9&6330&35.552&3.74&G5Iab/Ib&1\\
V397 Car&87072&2.063&0.001&49.6&0.6&6150&2.063&8.30&F8IB/II&1\\
RY Vel&89841&27.952&0.177&201.2&3.2&37433&28.136&8.40&F5Ib/II&1\\
VY Car&93203&18.852&0.033&160.4&2.3&5796&18.890&7.62&F7Iab/Ib&1\\
U Car&95109&38.609&0.315&247.6&3.3&5757&38.829&6.45&G3Ia&1\\
ER Car&97082&7.722&0.011&103.0&1.4&5914&7.719&6.82&G1Iab/Ib&1\\
IT Car&97485&7.524&0.008&68.1&1.0&5990&7.533&8.11&F8Iab/b&1\\
V419 Cen&100148&5.502&0.005&61.7&1.0&7716&5.507&8.18&F7II&1\\
S Mus&106111&9.689&0.012&108.3&1.4&19678&9.660&6.08&F6Ib&1\\
R Cru&107805&5.818&0.004&140.4&2.1&10769&5.826&6.90&F7Ib/II&1\\
BG Cru&108968&3.345&0.002&45.4&0.6&10853&3.343&5.49&F5III&1\\
AG Cru&110258&3.836&0.004&89.5&1.5&10829&3.837&8.23&F8Ib/II&1\\
R Mus&110311&7.529&0.010&196.7&2.1&19677&7.510&7.51&F7Ib&1\\
S Cru&112044&4.687&0.001&146.9&2.0&10770&4.690&6.73&F7Ib/II&1\\
V659 Cen&117399&5.629&0.007&59.4&0.8&7956&5.623&6.65&F6/F7Ib&1\\
XX Cen&118769&10.938&0.021&171.5&2.4&7346&10.953&7.83&F7/F8II&1\\
V381 Cen&120400&5.080&0.002&151.2&2.1&7678&5.079&7.68&F8Ib/II&1\\
V Cen&127297&5.482&0.003&132.6&2.1&7302&5.494&6.80&F5Ia&1\\
AV Cir&130233&3.066&0.002&74.6&0.8&9474&3.065&7.44&F7II&1\\
AX Cir&130701&5.279&0.015&76.1&1.1&9476&5.273&5.94&F8II+A/F&1\\
--&132247\tablenotemark{a}&2.123&0.008&8.7&0.2&412415&2.122&8.10&A0IV&3\\
R TrA&135592&3.392&0.002&129.8&1.3&36665&3.389&6.70&F7Ib/II&1\\
--&136633\tablenotemark{a}&6.118&0.009&28.0&0.5&412524&6.125&8.21&B3V&1\\
LR TrA&137626&2.429&0.001&32.3&0.4&36930&2.428&7.79&F8II&1\\
S TrA&142941&6.324&0.006&162.3&1.8&36666&6.324&6.45&F8II&1\\
U TrA&143999&2.567&0.002&151.4&1.8&36668&2.568&7.92&F8Ib/II&1\\
S Nor&146323&9.754&0.018&84.1&1.0&19962&9.754&6.53&F8/G0Ib&1\\
RV Sco&153004&6.067&0.011&155.3&5.0&32830&6.061&7.16&G0Ib&2\\
V636 Sco&156979&6.803&0.010&93.5&1.5&33452&6.797&6.68&F7/F8Ib/II&3\\
V482 Sco&158443&4.529&0.007&106.0&3.5&33298&4.528&7.93&F8/G0II&2\\
V950 Sco&159654&3.378&0.001&67.0&1.1&33766&3.380&7.27&F5Ib&3\\
X Sgr&161592&7.018&0.003&278.0&8.6&27707&7.013&4.56&F7II&2\\
RY Sco&162102&20.063&0.007&172.8&5.1&32833&20.323&8.18&F6Ib&2\\
W Sgr&164975&7.597&0.001&172.4&6.9&27706&7.595&4.70&G0Ib/II&2\\
kap Pav&174694&9.031&0.008&180.9&2.1&25119&9.083&4.36&F5Ib-II:&1\\
XY Car&308149&12.430&0.055&5.5&0.1&5803&12.434&6.97&A9Ib-II&1
\enddata
\tablenotetext{a}{Reclassified in this work, see \S4.1.} 
\tablerefs{(1) \citet{Houk75}, (2) \citet{Houk82}, (3) \citet{Houk78}.}
\end{deluxetable*}

A total of 47 previously classified Cepheid variables detected with bRing have well-defined periods in the VSX catalogue. 
These are tabulated in Table \ref{cepht}, which includes identification information about each star, the primary bRing period and amplitude, and the reported VSX period. 
This structure is used for tables throughout this paper. 
The primary frequencies recovered by bRing agreed with all of the fundamental modes reported in the VSX catalogue. 
A future study of Cepheids in bRing could yield fainter frequency modes present in their power spectrum and help identify the Bla\v{z}hko effect if present \citep{Blazko07}. 
Based on their CMD position (Figure \ref{cmd}), two of the stars are unusual for Cepheids; we reclassify them:


 {\it HD 132247} (ASAS J145955-4957.9) is a A0IV star \citep{Houk78} classified in VSX as both a first-overtone classical Cepheid and an $\alpha^{2}$ Canum Venaticorum (ACV) \citep{Sitek12}. This is a poorly studied star that does show an 8 mmag pulsation at a period of 2.123 $d$. There are other modes present in the star's periodogram, however, nothing is indicative of it being a classical Cepheid in addition to its spectral classification. Although its period could indicate this is an ACV variable, a lack of spectral observations to identify chemical peculiarities and spectral line intensity variations make it challenging to unambiguously classify. 
 
 One possible variable classification is a $\delta$ Scuti. Although the luminosity class of the star suggests it lies beyond the blue edge of the instability strip \citep{Breger98}, its position in Figure \ref{cmd} ($(B-V)_0$ = 0.13, $M_V$ = 1.40) is on top of the other $\delta$ Scutis in this study. In addition, $\delta$ Scutis have been shown to exist blueward of this theoretical limit \citep{Bowman18,Mellon19}. Therefore, the classical Cepheid designation should be removed. The ACV designation should also be changed due to lack of a detailed spectral study. It is reasonable to suggest that this star is actually a $\delta$ Scuti based on its CMD position and multiple pulsation modes present in its periodogram.
 

 {\it HD 136633} (ASAS J152459-6156.7) is a B3V star \citep{Houk75} classified as a fundamental-mode classical Cepheid in VSX \citep[from][]{Sitek12}. The periodogram does reveal multiple modes present, however, the B3V spectral classification means this star astrophysically is unlikely to be a classical Cepheid, however it is more likely to be a $\beta$ Cephei (BCEP) or slowly pulsating B-type (SPB) star.
This agrees with its position on the CMD (Figure \ref{cmd}: $(B-V)_0$ = -0.12, $M_V$ = -2.00).
 The modes present in this star appear to better fit the description of an SPB and should be reclassified as such \citep{Miglio07,DeCat07}.
  
\subsection{Eclipsing Binaries and the O'Connell effect} 

We detected 120 eclipsing binaries (EBs) in the bRing dataset.
For most of these EBs, the periodogram revealed the half period (the phase-folded light curve showed the primary and secondary eclipses overlapping) as the dominating sinusoidal component.
When a potential EB was found, the periodogram was re-scanned in a window around double the original period to find the true period. 
For a few EBs, this was not true and the correct period was searched for manually.  
The known EBs were discussed in \S \ref{sec:knownebs} and tabulated in Table \ref{ebst} similarly to the Cepheids. 
Periods from \citet{Rimoldini12} were used in place of missing VSX periods where available.  
Three new EBs are discussed in \S\ref{newEBs} and summarized in Table \ref{newebs}. 
Eighteen of the bRing EB light curves also showed evidence of the O'Connell effect \citep{OConnell51}, and these are discussed in \S \ref{sec:oconnell} and their parameters are summarized in Table \ref{oconn}.

\startlongtable
\begin{deluxetable*}{c|c|c|c|c|c|c|c|c|c|c}\tablecaption{Previously Classified Eclipsing Binaries Detected With bRing \label{ebst}}
\tablehead{\colhead{Name} &\colhead{HD} & \colhead{P} & \colhead{$\sigma_{P}$} & \colhead{A} & \colhead{$\sigma_{A}$} &\colhead{VSX ID} & \colhead{P$_{VSX}$} & \colhead{$V$} & \colhead{SpT} & \colhead{Ref}\\
\colhead{...} &\colhead{...} & \colhead{d} & \colhead{d} & \colhead{mmag}& \colhead{mmag}& \colhead{...} & \colhead{d} & \colhead{mag} & \colhead{...}& \colhead{...}}
\startdata
zet Phe&6882&1.66985&1e-04&36.7&0.5&26329&1.66978&3.98&B6V+B0V&1\\
--&16589&6.33296&2e-04&5.4&0.2&53991&0.82414\tablenotemark{a}&6.48&F6V&2\\
CN Hyi&17653&0.45609&2e-04&68.6&0.6&16473&0.45611&6.67&F6V&1\\
WZ Hor&17755&0.72886&1e-04&48.0&0.6&15947&0.72885&8.06&F3/F5V&1\\
VY Ret&21765&14.21605&5e-04&3.3&0.2&39786&14.21605&7.89&F5V&1\\
RZ Cae&29087&2.48712&5e-04&10.2&0.4&4529&2.48696&7.83&A4V&2\\
AN Dor&31407&2.03274&1e-04&13.2&0.3&13656&2.03268&7.67&B2/B3V&1\\
AR Dor&34349&2.95130&8e-05&4.7&0.1&13660&2.95206&7.03&F5V&1\\
UX Men&37513&4.18110&1e-03&25.5&0.8&18670&4.18110&8.25&F8V&1\\
TY Men&37909&0.46166&1e-04&99.0&1.0&18665&0.46167&8.26&A3/A4V&1\\
del Pic&42933&1.67248&2e-04&46.9&0.6&26396&1.67254&4.72&B0.5IV&1\\
V360 Pup&52993&1.12803&2e-04&10.7&0.3&26962&1.29644&6.57&ApSi&2\\
V361 Pup&54579&0.23661&5e-04&49.8&2.7&26963&0.36737&8.04&G0V&3\\
FF CMa&55173&1.21332&4e-04&67.2&3.2&5323&1.21337&7.48&B3/5V(p)&2\\
V452 Car&56146&2.11033&1e-05&23.2&0.4&6205&1.05502&8.10&B8IV&1\\
--&56910&1.83724&2e-05&5.7&0.2&55845&0.94929\tablenotemark{a}&6.84&A2/3mA4-A7&1\\
V376 Pup&60559&3.88333&2e-04&3.8&0.2&26978&1.94270&6.25&B8IV(p Si)&2\\
V454 Car&60649&0.98049&1e-04&32.6&0.5&6207&0.98042&6.99&B4/B5V&1\\
V455 Car&61644&5.13038&2e-04&14.4&0.3&6208&5.13300&8.40&B5/B6IV&1\\
V606 Car&63203&12.31530&2e-03&9.5&0.4&42349&12.31920&8.31&B8/B9III&1\\
V397 Pup&63786&3.00402&2e-04&3.4&0.2&26999&3.00445&5.93&B9V&2\\
QZ Pup&64503&1.11207&2e-04&8.4&0.2&26936&1.11203&4.48&B2V&2\\
V Pup&65818&1.45441&2e-04&115.9&1.5&26607&1.45449&4.49&B1Vp+B2&4\\
--&66623&0.85182&1e-04&7.6&0.6&250227&0.42573&8.11&F7V&2\\
V462 Car&66768&1.10561&6e-05&30.7&0.4&6215&1.10569&6.71&B3V(n)&1\\
V431 Pup&69882&9.34999&1e-04&10.9&0.3&27033&9.35928&7.18&B1III:&5\\
--&70999&1.99520&5e-03&14.1&0.6&358580&2.99250&8.04&B3III&2\\
HR 3322&71302&4.93500&2e-04&4.4&0.2&27040&4.93500&5.97&B3V&5\\
NO Pup&71487&0.77183&2e-03&10.3&0.3&26892&1.25689&6.50&B9IV/V&2\\
XY Pyx&71801&0.92254&4e-04&10.1&0.4&27231&0.92254&5.74&B2V&2\\
X Car&72698&0.54132&1e-04&46.5&0.8&5760&1.08263&8.06&A0Vn&1\\
FY Vel&72754&33.88620&5e-04&39.2&0.6&37604&33.72000&6.89&B2Iape&5\\
V470 Car&72878&2.16177&2e-04&19.9&0.4&6223&2.16178&7.47&B9IV&1\\
V454 Vel&73699&1.13484&2e-04&16.1&0.4&272444&1.13492&7.58&B3V&2\\
NX Vel&73882&2.91834&3e-04&6.0&0.3&37715&2.91988&7.26&O8V:&6\\
RS Cha&75747&1.66999&1e-04&51.3&0.6&9248&1.66987&6.08&A7V&1\\
CV Vel&77464&6.89145&3e-04&2.1&0.1&37538&6.88949&6.70&B2V+B2V&5\\
GP Vel&77581&8.97155&2e-04&10.8&0.4&37614&8.964357&6.91&B0.5Ib&5\\
PQ Vel&78165&22.2632&1e-03&4.5&0.2&37731&22.2632&7.61&A2/3III(m)&5\\
V476 Car&78763&1.28135&2e-03&15.2&0.3&6229&1.28143&8.30&B7Vn&1\\
S Vel&82829&5.93101&2e-03&23.1&0.4&37418&5.93365&7.80&A5Ve+K5IIIe&7\\
IP Vel&84400&3.43679&1e-04&23.3&0.4&37649&3.43789&6.16&B6V&5\\
V486 Car&84416&1.09378&1e-04&32.7&0.4&6239&1.09389&6.32&A0V&1\\
KN Vel&85037&2.72327&1e-04&7.9&0.2&37663&2.72290&6.52&A2IV(m)&5\\
QX Vel&85185&0.87811&2e-04&37.8&0.6&37748&0.87807&8.00&A0V&5\\
QX Car&86118&4.47804&1e-04&14.0&0.3&6085&4.47804&6.66&B3V+B3V&1\\
V367 Car&86441&5.71172&2e-04&13.6&0.3&6120&5.73000&7.52&B6V&1\\
V341 Vel&89611&14.73000&9e-04&2.8&0.3&37757&14.73000&7.96&A0IV&5\\
V435 Vel&90000&10.49500&7e-04&4.7&0.2&37761&10.49500&7.56&B3V&5\\
--&90941&7.56760&4e-04&1.1&0.2&411431&7.56470&7.87&B4IV&5\\
CC Ant&91519&2.44594&6e-05&19.7&0.6&172655&2.44514&7.70&A8III&2\\
V661 Car&93130&0.39875&1e-04&14.8&0.4&56932&23.9438&8.08&O6III&6\\
RZ Cha&93486&2.83200&2e-04&24.8&0.4&9255&2.83208&8.08&F5V+F5&1\\
V356 Vel&93668&1.76804&2e-04&14.6&0.3&37772&1.76791&6.74&A0V&5\\
V772 Car&94924&0.88419&2e-04&27.4&0.4&172663&0.88417&8.01&A1V&1\\
V529 Car&95993&4.74574&2e-04&33.8&0.6&6282&4.74461&8.18&B8V&1\\
TU Mus&100213&1.38710&1e-04&101.9&1.3&19704&1.38728&8.40&O8(+O8)&6\\
V1101 Cen&102682&5.03350&2e-04&25.0&0.6&43976&5.03230&8.23&F5V&5\\
LZ Cen&102893&2.75772&1e-04&88.6&1.2&7571&2.75772&8.24&B2III&1\\
V788 Cen&105509&4.96697&1e-03&1.9&0.1&8085&4.96638&5.74&A3III&5\\
V831 Cen&114529&0.32142&5e-03&5.0&0.1&8128&0.64252&4.58&B8V&1\\
V964 Cen&115823&1.54308&1e-04&6.5&0.1&8261&1.54259&5.45&B6V&5\\
V979 Cen&119888&2.56882&7e-05&18.0&0.3&8276&2.56841&7.84&B8II&1\\
V1294 Cen&121291&1.16556&2e-04&25.4&0.5&45116&1.16553&7.89&A0Vn+K2(III)&5\\
AT Cir&122314&3.25748&3e-04&9.8&0.2&9472&3.25749&7.62&A5IV/Vs&1\\
V992 Cen&122844&1.21168&1e-04&16.4&0.3&8289&1.21156&6.20&A5III/IV&1\\
--&123720&0.86872&5e-05&16.4&0.3&58490&0.86880&7.75&A4V&1\\
V716 Cen&124195&1.49024&2e-04&32.9&0.5&8013&1.49010&6.09&B5V&8\\
RR Cen&124689&0.60570&1e-04&71.8&1.1&7307&0.60569&7.46&A9/F0V&1\\
--&129094&0.39881&2e-03&27.1&0.8&98784&0.74422&8.37&F7V&1\\
QZ Lup&131638&1.13658&1e-04&15.6&0.3&45479&1.13655&8.32&B9V&5\\
HR Lup&133880&1.75470&1e-04&11.9&0.3&17811&0.87748&5.76&B8IVSi&9\\
del Cir&135240&3.90445&1e-04&19.2&0.3&9529&3.90248&5.07&O8.5V&6\\
GG Lup&135876&1.84961&1e-03&4.9&0.3&17783&1.84961&5.59&B9V&5\\
MP TrA&143028&2.07017&3e-04&8.3&0.2&36942&2.06972&7.80&B7Ib/II&1\\
V399 Nor&147170&3.19301&3e-04&13.3&0.3&59034&3.19288&8.21&F6/F7V&3\\
V760 Sco&147683&1.73074&2e-04&15.1&0.6&33576&1.73090&7.05&B4V&2\\
OT Aps&148891&2.42603&9e-05&6.7&0.2&832&2.42660&8.00&B9.5IV&1\\
V1288 Sco&149450&1.10896&1e-05&32.0&0.6&46471&1.10890&8.23&B3III&5\\
V882 Ara&149668&20.96590&9e-05&4.3&0.2&59156&20.96590&7.61&A2IV&1\\
R Ara&149715&8.85166&2e-03&25.7&0.8&2804&4.42522&8.33&K0III&1\\
V954 Sco&149779&1.26883&1e-04&48.9&0.8&33770&1.26859&7.57&B2IV&5\\
V878 Ara&151475&0.77053&6e-05&46.0&0.7&136724&0.77046&8.05&B3II/III&5\\
V1290 Sco&151564&4.49267&2e-04&6.0&0.3&59217&4.49244&7.98&O9.5IV&5\\
HR 6247&151890&1.44647&1e-04&44.8&0.8&34007&1.44627&2.99&B1.5IV+B&10\\
V1295 Sco&152333&2.15767&3e-04&32.4&0.6&59262&2.15767&8.07&B1-2Ib-II&5\\
V861 Sco&152667&7.85382&1e-04&38.4&0.8&33677&7.84818&6.18&B0.5Ia&2\\
V883 Sco&152901&4.34113&1e-04&23.6&0.5&33699&4.34119&7.39&B2.5Vn&11\\
V836 Ara&153140&7.04075&2e-04&24.8&0.4&3639&7.03418&7.51&B1II&5\\
V616 Ara&154339&4.99671&6e-05&52.4&0.8&3419&4.99525&8.26&B3II/III&5\\
FV Sco&155550&5.72861&2e-04&37.9&1.4&33001&5.72790&8.07&B4IV&2\\
V1012 Sco&155775&1.51531&2e-04&12.0&0.2&33828&1.51548&6.72&B1V&6\\
V499 Sco&158155&2.33216&2e-04&67.2&2.5&33315&2.33330&8.29&B1III&2\\
V1081 Sco&158186&2.51419&1e-04&10.5&0.6&33897&2.51374&7.00&O9.5V(n)&6\\
V535 Ara&159441&0.31466&2e-04&30.2&0.4&3338&0.62930&7.36&A8V&1\\
V539 Ara&161783&3.16836&3e-04&19.6&0.3&3342&3.16909&5.70&B2V+B3V&1\\
V453 Sco&163181&12.00201&7e-05&82.7&2.4&33269&12.00597&6.60&O9.5Ia/ab&2\\
V1647 Sgr&163708&3.28277&1e-04&24.5&1.0&29347&3.28279&7.06&A3III&2\\
V2509 Sgr&167231&1.84197&4e-04&34.9&1.4&30209&1.08697&7.41&A0IV&2\\
V681 CrA&171577&4.32961&2e-04&2.8&0.2&10552&4.32788&7.74&B9V&5\\
V362 Pav&173344&2.74826&7e-05&5.3&0.1&25082&2.74844&7.39&A2mA5-A9&1\\
V363 Pav&174139&1.19491&1e-04&27.7&0.4&25083&1.19497&8.17&B9/B9.5V&1\\
V4407 Sgr&174632&1.45165&1e-04&19.1&0.8&32107&1.45174&6.64&B7/B8IV&8\\
--&177776&1.65022&7e-05&12.7&0.3&414518&1.65006&8.12&B9.5Vn&1\\
V4089 Sgr&184035&4.62891&3e-04&9.5&0.3&31789&4.62988&5.90&A5IV-III&5\\
HO Tel&187418&1.61294&2e-04&53.9&0.8&36458&1.61310&8.30&A7III(m)&5\\
V4437 Sgr&193174&1.13654&3e-04&36.0&1.3&32137&1.13662&7.24&A9IV/V&2\\
V386 Pav&198736&0.55187&2e-04&31.2&0.4&25106&0.55184&8.34&A9V&1\\
DE Mic&200670&0.20535&2e-04&16.6&0.5&137558&0.41069&7.80&F6/7V&2\\
BR Ind&201427&1.78553&2e-04&6.7&0.2&16577&0.89277&7.09&F8V&3\\
--&203244\tablenotemark{b}&12.77751&7e-03&9.0&0.1&64006&833.29734\tablenotemark{a}&6.98&G5V&3\\
CH Ind&204370&5.94788&2e-03&12.9&0.3&137591&5.95320&7.52&A9V&5\\
--&205877&7.68402&3e-04&7.4&0.1&64150&3.83266\tablenotemark{a}&6.20&F7III&5\\
CP Gru&208614&2.08577&3e-04&29.1&0.4&14785&2.08615&7.72&A5V&5\\
DV Gru&210572&9.61553&1e-03&2.9&0.1&64287&4.81803&7.72&F8V&1\\
DK Tuc&212661&5.33386&4e-04&5.2&0.2&37084&5.33793&6.90&A1mA5-F0&1\\
DP Gru&220633&3.80231&3e-04&9.5&0.3&14807&3.80350&8.29&F5/F6V&5
\enddata
\tablenotetext{a}{\citet{Rimoldini12}}
\tablenotemark{b}{Reclassified}
\tablerefs{
(1) \citet{Houk75}, 
(2) \citet{Houk82}, 
(3) \citet{Torres06}, 
(4) \citet{Hiltner69}, 
(5) \citet{Houk78}, 
(6) \citet{Sota14}, 
(7) \citet{Sahade52},
(8) \citet{Hube70},
(9) \citet{Buscombe69}, 
(10) \citet{Levato75},
(11) \citet{Garrison77}.
}
\end{deluxetable*}


\subsubsection{Previously Identified Eclipsing Binaries}\label{sec:knownebs}

For the 117 variable stars previously classified as EBs, the measured periods were compared to the values listed in the VSX catalogue.
The periods we measured with bRing agreed with 95 ($\sim$81\%) of the periods reported in VSX within 1\%.
There were 9 stars ($\sim$7\%) whose bRing periods were double the VSX period, but for which we confirmed the bRing periods by visual inspection of the phase-folded light curve (DV Gru, BR Ind, V452 Car, HD 66623, HD 205877, HD 56910, V376 Pup, HR Lup, R Ara).
For 8 stars ($\sim$7\%) a different period was detected that reproduces the eclipse structure whereas the VSX period does not (V361 Pup, HD 70999, V360 Pup, V2509 Sgr, V661 Car, HD 16589, NO Pup, HD 203244). 
There were also 5 stars ($\sim$5\%) whose bRing periods recovered the eclipsing structure at half the reported VSX period (DE Mic, HD 129094, X Car, V535 Ara, V831 Cen).
We further discuss a couple notable cases: HD 70999 and HD 203244.

\startlongtable
\begin{deluxetable*}{c|c|c|c|c|c|c|c|c}
\tablecaption{New Eclipsing Binaries Detected With bRing \label{newebs}}
\tablehead{\colhead{HD} & \colhead{P} & \colhead{$\sigma_{P}$} & \colhead{Pri} & \colhead{Sec} &\colhead{VSX ID} & \colhead{$V$} & \colhead{SpT} & \colhead{Ref}\\
\colhead{...} & \colhead{d} & \colhead{d} & \colhead{mag}& \colhead{mag}& \colhead{...} &\colhead{mag} & \colhead{...}& \colhead{...}}
\startdata
77669&7.70766&5e-04&0.180&0.15&--&8.10&B9III/IV&1\\
142049&13.22062&4e-05&0.050&0.045&45942&5.85&G5II/III+A3&2\\
155781&13.08670&6e-05&0.100&0.080&--&7.42&A3IV/V&2\\
\enddata
\tablerefs{(1) \citet{Houk78}, (2) \citet{Houk75}.}
\end{deluxetable*}


{\it HD 70999}: Unfortunately, the VSX period for this star was near one of the strong, low-frequency bRing systematic false positives (see \S \ref{FP}). 
The light curve also seemed to be missing the eclipses for $\sim$25\% of the observations. 
When phase-folding on the VSX period of 2.99250 d, the phase-folded light curve showed the dip broken up into three segments with no clear eclipse structure. 
When bRing data were phase-folded to 1.9952 d, 2 dips were recovered, but no clear eclipse structure was seen. 
Due to this lack of data in bRing, the period for this eclipsing system was not accurately determined.

{\it HD 203244} was classified as an Algol eclipsing binary (EA) in both the VSX catalog and \citet{Rimoldini12}, with the latter reporting an unusually long period of 833.29734 $d$. 
We detect in the bRing photometry a very strong period at 12.77751 $d$, however the phase-folded light curve at this period is shallow and does not show a secondary eclipse as expected from an EA. 
Phase-folding on the half or double period did not reveal additional structure. 
HD 203244 is most likely an ellipsoidal variable based on the period and shape of the phase-folded light curve.

\subsubsection{New Eclipsing Binaries \label{newEBs}} 
\begin{figure*}
\begin{center}
\gridline{\fig{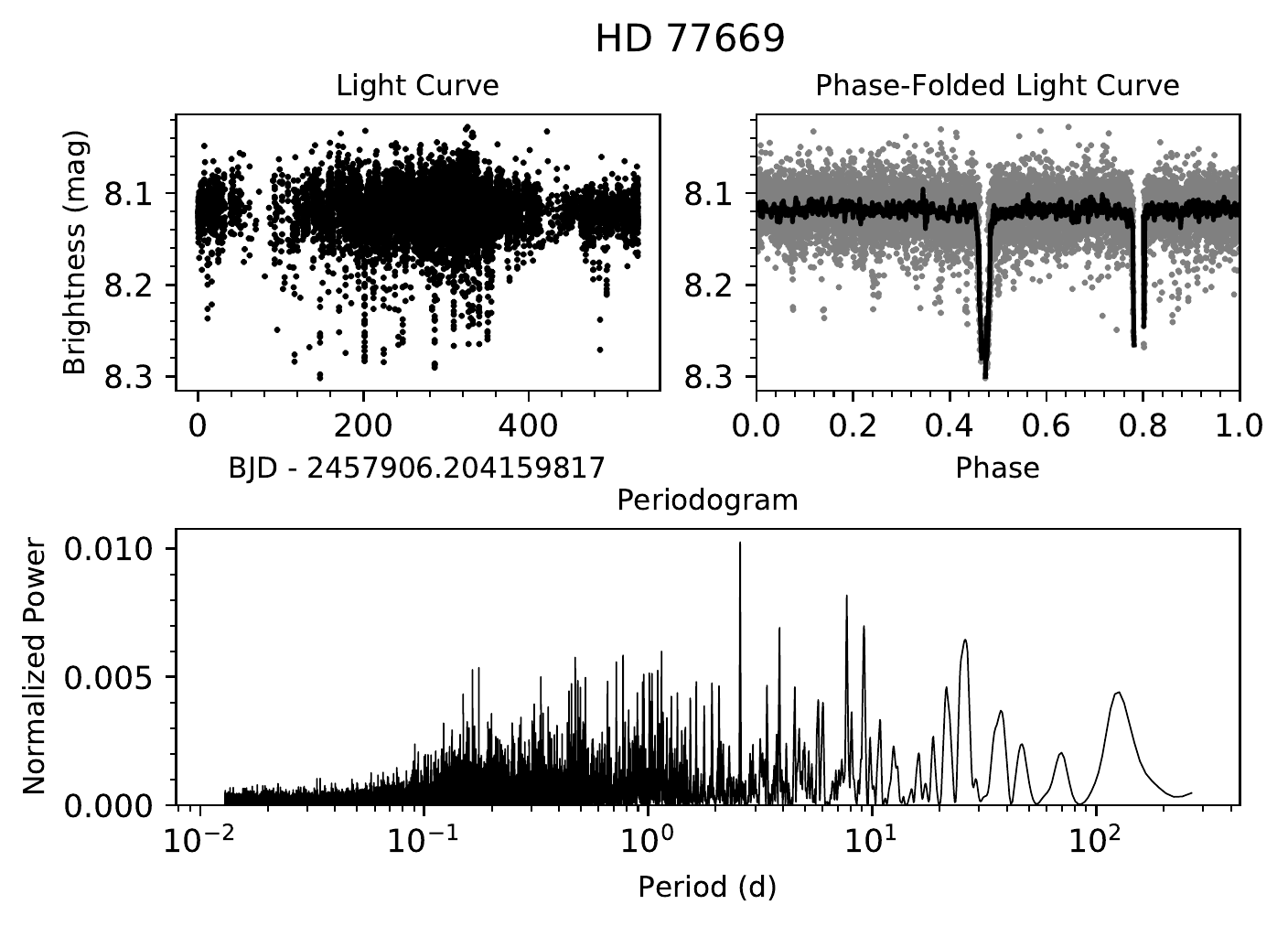}{0.3\textwidth}{(a)}
          \fig{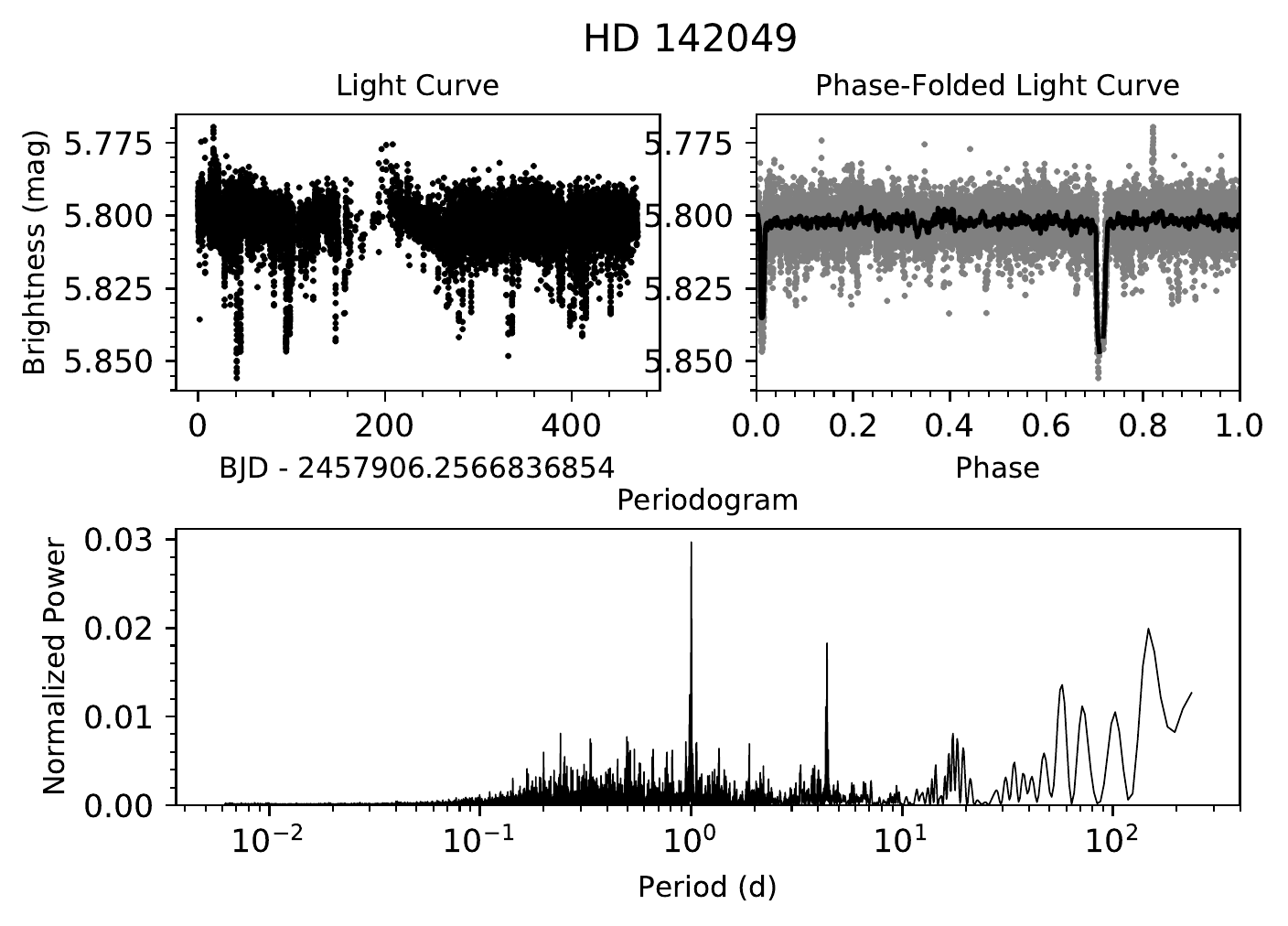}{0.3\textwidth}{(b)}
          \fig{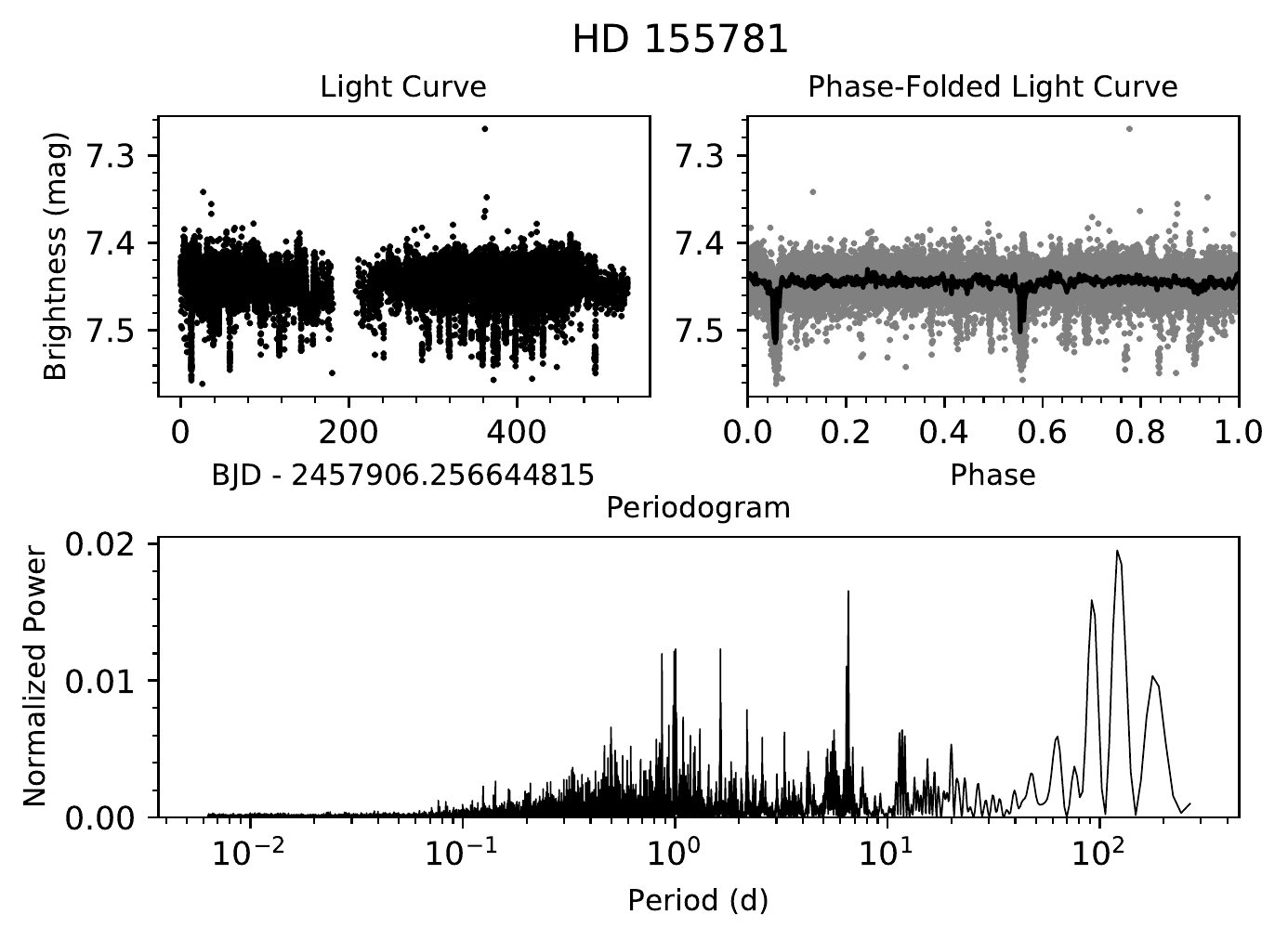}{0.3\textwidth}{(c)}}
\caption{The light curves, periodograms, and phase-folded light curves of the three new eclipsing binaries detected in this survey.}
\label{3ebsplot}
\end{center}
\end{figure*}
Three new eclipsing binaries were identified and their phase-folded light curves are shown in Figure \ref{3ebsplot}. To better identify the orbital periods for these new eclipsing binaries, a BLS routine adopted from other works in this group was used \citep[e.g.,][]{TalensM1,Dorval19}.

{\it HD 77669}: This is a B9III/IV star \citep{Houk78}
with $V$ magnitude 8.11 \citep{ESA97} 
and parallax $\varpi$ = 1.9553\,$\pm$\,0.0514 mas,
corresponding to distance $d$ = 511\,$\pm$\,13 pc \citep[Gaia DR2 5331845690580305920;][]{GaiaDR2}.  
We detect a strong period of 7.70766 $d$ with a primary eclipse depth of 0.18 mag and a secondary eclipse depth of 0.15 mag. 
The star is fairly faint for bRing, and transits are seen
of depth $>$0.15 mag depth, however we can not accurately
measure the depths of the primary or secondary eclipses.

{\it HD 142049 (HR 5900)}: HD 142049 is catalogued in VSX as a suspected variable (NSV 7318). 
The Washington Double Star catalog \citep{Mason19}
reports HD 142049 as a 4\arcsec.8  binary with
$V$ magnitudes of 5.91 and 8.36, and the common motion  and parallax of the pair is obvious in Gaia DR2 \citep[Gaia DR2 5833110434699732352 and Gaia DR2 5833110434673386240;][]{GaiaDR2}.
The Gaia DR2 parallaxes are 18.1136\,$\pm$\,0.0678 mas 
and 18.1238\,$\pm$\,0.0445 mas for A and B, respectively, 
showing the resolved pair to be at distance 55.2 pc. 
The spectral types of the components are a matter of some contention, with \citet{Houk75} reporting components of 
type G5II/III and A3, noting that there is "{\it slight possibility there is a Am or Fm star component}", and \citet{Corbally84} 
reporting types of kA3hF3mF4 for the primary and F9.5V
for the secondary. 
The bRing data show that the unresolved light from the system is consistent with a grazing eclipsing binary with period 13.22062 day, with primary eclipse depth 0.050 mag and secondary eclipse depth of 0.045 mag. 
The binary must have a fairly eccentric orbit as the eclipses are only 0.3 phase apart.  

{\it HD 155781}: This $V$ = 7.43 star has spectral type A3IV/V \citep{Houk75,ESA97} and parallax 3.9861\,$\pm$\,0.0561 mas, corresponding to distance $d$ = 250.9\,$\pm$\,3.5 pc \citep[Gaia DR2 5913908252773468928;][]{GaiaDR2}.
We detect a strong signal at a period of 13.08670 days which appears to correspond to the orbital period of an eclipsing binary with primary dips of 0.10 mag and secondary dips of 0.08 mag.

\subsubsection{The O'Connell Effect}\label{sec:oconnell} 
This survey searched for evidence of the O'Connell effect in all of the W UMa and $\beta$ Lyr EBs in this data set. In W UMa and $\beta$ Lyr EBs, the O'Connell effect is observed as an asymmetry in the maximum brightness in between the primary and secondary eclipses, i.e., the maximum before the primary eclipse is fainter than the maximum before the secondary eclipse \citep{OConnell51}. The underlying physical mechanism is not well understood \citep[plausible explanations include surface features and Doppler beaming:][]{Wilsey09,dasilva14} though several examples have been detected \citep{Pribulla03,Pribulla11,Burggraaff18}. 

The O'Connell effect was detected in 18 of the bRing EBs, which have been tabulated in Table \ref{oconn}. The differences between the maxima was considered significant if it exceeded 3$\sigma_A$ where $\sigma_A$ is the uncertainty in the amplitude of the EB. Only two of the EBs in this table (TY Men \citep{Nagy85,Pribulla11} and TU Mus \citep{Terrell03} have been noted in the literature as having evidence of asymmetry in their light curves. The other 16 have likely been missed due to the faint effect observed in bRing and possible variability of the effect \citep{Wilsey09} masking the asymmetry in previous studies.
\startlongtable
\begin{deluxetable}{c|c|c|c|c|c}
\tablecaption{Eclipsing Binaries Showing the O'Connell Effect Detected With bRing \label{oconn}}
\tablehead{\colhead{Name} &\colhead{A} & \colhead{$\sigma_{A}$} & \colhead{Max 1} & \colhead{Max 2} & \colhead{$\Delta$m}\\
\colhead{...} & \colhead{mmag} & \colhead{mmag} & \colhead{(mag)}& \colhead{(mag)}& \colhead{mmag} }
\startdata
del Pic&46.9&0.6&4.6275&4.6301&2.6\\
V462 Car&30.7&0.4&6.5196&6.5120&7.6\\
TY Men&99.0&1.0&8.0567&8.0925&35.8\\
V535 Ara&30.2&0.4&7.3529&7.3551&2.2\\
V954 Sco&48.9&0.8&7.4828&7.4933&10.5\\
V772 Car&27.4&0.4&8.0544&8.0580&3.6\\
QZ Pup&8.4&0.2&4.4553&4.4584&3.1\\
DE Mic&16.6&0.5&7.7230&7.7323&9.3\\
HD 123720&16.4&0.3&7.7724&7.7772&4.8\\
LZ Cen&88.6&1.2&8.3148&8.3288&14.0\\
V979 Cen&18.0&0.3&7.4619&7.4654&3.5\\
V1012 Sco&12.0&0.2&6.6502&6.6525&2.3\\
V716 Cen&32.9&0.5&6.0115&6.0072&4.3\\
QX Vel&37.8&0.6&7.8886&7.8824&6.2\\
HR 6247&44.8&0.8&2.8767&2.8705&6.2\\
V470 Car&19.9&0.4&7.2323&7.2298&2.5\\
TU Mus&101.9&1.3&8.2469&8.2266&20.4\\
RR Cen&71.8&1.1&7.2808&7.2752&5.6
\enddata
\end{deluxetable}

\subsection{$\delta$ Scuti Variables} \label{sec:ds}

We detected 66 $\delta$ Scuti variables in the bRing dataset, 26 of which are candidates which had not been previously reported as detected pulsators. 
For the 40 previously known $\delta$ Scutis, we report only the strongest frequency in the bRing light curve.
The previously published frequencies for the $\delta$ Scuti variables in Table \ref{deltas} are from VSX by default,  however if one was not listed in VSX, we cite additional sources \citep{Rimoldini12,Rodriguez00,Mellon19}.
The reported periods for 22 (55\%) of the $\delta$ Scuti variables in Table \ref{deltas} do not match those reported in previous studies.
All of the $\delta$ Scutis are very tightly bound with their positions on the CMD (Figure \ref{cmd}; this is useful for confirming the new $\delta$ Scutis candidates by inspection.
The bright $\delta$ Scuti variable $\beta$ Pictoris itself was not included in this work since the bRing data for $\beta$ Pictoris were recently published and analyzed in \citet{Zwintz19}.
The mismatches may be due to a variety of reasons including aliasing or the presence of multiple modes, however, the star $\theta$ Tuc had an additional feature in its periodogram that is not $\delta$ Scuti in nature.

{\it HD 3112 ($\theta$ Tuc)}: $\theta$ Tuc is a well-studied $\delta$ Scuti that is also a well-studied binary system \citep[e.g.,][]{Cousins71,Stobie76,Kurtz80,Bos94,Sterken97,DeMey98}. The primary pulsation reported in the VSX catalogue is 20.28068 $d^{-1}$, which agrees with prior observations \citep{Cousins71,Stobie76,Kurtz80,Liakos17}. However, the dominant period detected by bRing is 0.28165 $d^{-1}$, which is reported in Table \ref{deltas}. This pulsation has been previously identified as orbital motion associated with the binary nature of the system \citep[0.281 $d^{-1}$:][]{Sterken97,DeMey98}. A search of the bRing periodogram around the expected $\delta$ Scuti frequencies recovers a primary $\delta$ Scuti frequency of 17.06312 $d^{-1}$.

The 26 new candidate $\delta$ Scuti variables all had faint primary pulsation amplitudes of $<$10.5 mmag, with the exception of HD 216743, and they all had brightnesses in the range $V$ $\simeq$ 6.5 -- 8.3, and are reported in Table \ref{newds} (showing the primary pulsation frequency as seen by bRing).
Most of the newly discovered $\delta$ Scuti variables in the bRing survey were in the faint end of the magnitude range for the instrument ($V$ $<$ 6.5); the two brighter candidates were HD 171819 ($V$ = 5.84) and HD 189951 ($V$ = 5.25) \citep{Kharchenko09}.
%
%
%
Further analysis of the frequencies detected with the bRing time series photometry for the previously discovered and newly discovered $\delta$ Scuti variables is encouraged and out of the scope of this work. 

The star HD 140566 (included in Table \ref{otherp}) was labelled as detached eclipsing binary by the VSX with a period 193.70000 $d$. bRing detected a much shorter period at 0.08783 $d$ (11.38563 $d^{-1}$). This star is poorly studied with no prior follow-up work attempting to confirm the nature of this variable. The bRing period and light curve are not indicative of an eclipsing system. 
The combination of the detected pulsation in the bRing light curve and the star's spectral type (A5IV) indicate that the star is likely to be a $\delta$ Scuti variable. This agrees with its position among other $\delta$ Scutis in the CMD from Figure \ref{cmd} ($(B-V)_0$ = 0.16, $M_V$ = 0.99). Therefore, this star is not an eclipsing system and is reclassified in this work as a candidate $\delta$ Scuti.

\startlongtable
\begin{deluxetable*}{c|c|c|c|c|c|c|c|c|c|c}\tablecaption{Previously Classified $\delta$ Scutis Detected With bRing \label{deltas}}
\tablehead{\colhead{Name} &\colhead{HD} & \colhead{$f$} & \colhead{$\sigma_{f}$} & \colhead{A} & \colhead{$\sigma_{A}$} &\colhead{VSX ID} & \colhead{$f_{VSX}$} & \colhead{$V$} & \colhead{SpT} & \colhead{Ref}\\
\colhead{...} &\colhead{...} & \colhead{d$^{-1}$} & \colhead{d$^{-1}$} & \colhead{mmag}& \colhead{mmag}& \colhead{...} & \colhead{d$^{-1}$} & \colhead{mag} & \colhead{...}& \colhead{...}}
\startdata
$\theta$ Tuc&3112&0.28165&2e-04&4.0&0.2&37102&20.28068&6.11&kA7hA7mF0(IV)&1\\
--&8351&14.13144&5e-05&3.7&0.2&53727&14.06695\tablenotemark{a}&6.70&A9V&2\\
BD Phe&11413&27.02703&4e-03&4.9&0.2&26294&25.21158\tablenotemark{a}&5.93&A1Va $\lambda$ Boo&3\\
--&12284&6.25500&7e-05&5.2&0.3&53855&6.20694&7.68&A9III&2\\
RX Cae&28837&8.27307&5e-04&6.5&0.3&4527&6.48925&7.01&F3/F5II&4\\
X Cae&32846&0.29564&1e-04&5.0&0.3&4518&0.27049&6.31&F2IV/V&2\\
YY Pic&39244&18.51316&3e-04&3.7&0.2&26385&9.73985&7.79&A7V&4\\
--&41846&10.47770&4e-05&5.3&0.2&--&10.33475\tablenotemark{a}&8.12&A6mA7-F0&5\\
--&46586&14.82017&5e-04&2.9&0.2&410727&14.82052&8.04&F0III&4\\
%
V638 Pup&58635&8.66695&5e-03&2.5&0.1&26970&8.66699&6.82&A8V&2\\
V393 Car&66260&14.1551&1e-04&6.8&0.2&6146&7.07741&7.47&A7III/IV&5\\
AI Vel&69213&11.59990&1e-04&49.0&0.9&37479&8.96265&6.56&A9IV-V&4\\
OX Vel&77347&12.60254&1e-04&11.5&0.2&37727&12.60255&7.58&A4mA7-A9&5\\
ER Cha&88278&14.27857&1e-04&3.9&0.1&9398&15.72376&7.31&A3/5III/IV&5\\
LW Vel&88824&12.58582&5e-05&3.8&0.1&37687&8.98093&5.27&F0Vn&1\\
--&90611&15.19447&4e-04&2.5&0.2&--&15.19498\tablenotemark{a}&6.55&F0IV/V&4\\
IW Vel&94985&10.14809&2e-04&4.3&0.1&37656&6.66666&5.90&A4V&4\\
V1023 Cen&102541&19.89813&5e-03&4.1&0.2&8320&20.00000\tablenotemark{b}&7.95&hF0VkA5mA5 $\lambda$ Boo&6\\
EE Cha&104036&33.86956&3e-04&3.4&0.1&9386&33.33333\tablenotemark{b}&6.73&A7V&5\\
--&111984&23.49741&5e-05&3.3&0.1&--&21.46347\tablenotemark{a}&7.28&A5V&4\\
V853 Cen&126859&16.30857&8e-04&2.4&0.3&8150&18.92013&6.97&A6V&5\\
IN Lup&142994&9.16564&6e-05&3.0&0.2&17824&7.87402\tablenotemark{b}&7.17&F2VkA3mA3 $\lambda$ Boo?&7\\
IO Lup&143232&13.40241&5e-04&3.0&0.1&17825&15.59193\tablenotemark{a}&6.67&kA7hA5mF2&8\\
V922 Sco&153747&23.80734&1e-03&2.7&0.2&33738&20.00000&7.40&hA7VmA0 $\lambda$ Boo&6\\
--&156623&71.14754&8e-04&3.4&0.1&--&71.14300\tablenotemark{c}&7.24&A1V PHL&6\\
--&157321&10.51587&2e-05&9.7&0.2&60284&10.51640&8.02&A9IV/V&5\\
V703 Sco&160589&6.66876&6e-04&25.2&1.4&33519&8.67922&7.85&F0V&6\\
V346 Pav&168740&12.57296&1e-04&2.6&0.1&25066&16.98244\tablenotemark{d}&6.12&A8VkA2mA2 $\lambda$ Boo&7\\
V353 Tel&173794&0.61578&3e-04&4.7&0.1&137348&0.31250&7.11&A3III/IV&4\\
QQ Tel&185139&8.41297&3e-04&4.3&0.1&36568&15.38462&6.26&F2IV&4\\
--&192316&33.0221&1e-03&3.4&0.2&--&22.80867\tablenotemark{a}&7.55&A8V&5\\
--&198592&26.34244&8e-04&2.7&0.3&--&21.59235\tablenotemark{a}&7.58&A3III&4\\
--&200475&1.11423&2e-05&9.5&0.2&305774&1.11499&7.82&A3mA5-A7&5\\
--&201292&14.7769&4e-05&5.2&0.2&--&25.80477\tablenotemark{a}&8.19&A3II&5\\
CK Ind&209295&1.12963&3e-04&25.3&0.3&137616&1.12934&7.32&A9/F0V&5\\
DR Gru&213669&14.01768&7e-05&7.6&0.2&137628&15.01502&7.41& F0VkA2.5mA2.5 $\lambda$ Boo&7\\
--&218090&1.81606&1e-04&6.3&0.1&305852&1.81590&8.13&F0V&5\\
--&219301&9.28928&3e-05&3.6&0.2&250337&9.28966\tablenotemark{a}&6.56&F0III&5\\
RS Gru&--&13.60412&6e-05&46.9&0.7&14680&6.80218&8.27&A9IV&4\\
HIP 35815&--&7.70897&5e-05&6.9&0.3&55889&10.83677\tablenotemark{a}&7.84&F0&9
\enddata
\tablenotetext{a}{\citet{Rimoldini12}} 
\tablenotetext{b}{\citet{Rodriguez00}} 
\tablenotetext{c}{\citet{Mellon19}}
\tablenotetext{d}{\citet{Paunzen98}}
\tablerefs{
(1) \citet{GrayGarrison89},
(2) \citet{Houk82},
(3) \citet{GrayGarrison87},
(4) \citet{Houk78},
(5) \citet{Houk75},
(6) \citet{Paunzen01},
(7) \citet{Gray17},
(8) \citet{Paunzen96},
(9) \citet{Boss37}.
}
\end{deluxetable*}
\startlongtable
\begin{deluxetable*}{c|c|c|c|c|c|c|c|c}
\tablecaption{New Candidate $\delta$ Scutis Detected With bRing \label{newds}}
\tablehead{\colhead{HD} & \colhead{$f$} & \colhead{$\sigma_{f}$} & \colhead{A} & \colhead{$\sigma_{A}$} &\colhead{VSX ID}  & \colhead{$V$} & \colhead{SpT} & \colhead{Ref}\\
\colhead{...} & \colhead{d$^{-1}$} & \colhead{d$^{-1}$} & \colhead{mmag} &\colhead{mmag}& \colhead{...} & \colhead{mag} & \colhead{...}& \colhead{...}}
\startdata
3463&10.50785&5e-04&3.7&0.2&--&8.04&A6/8V&1\\
20232&45.45826&2e-04&2.1&0.2&--&6.88&A2/A3III/IV&2\\
25860&15.34882&4e-04&2.0&0.1&--&6.62&A4/A5IV&1\\
43898&19.31816&2e-04&2.6&0.3&--&7.87&A8/A9V&2\\
46978&16.6692&2e-03&4.4&0.2&--&8.16&A6V&1\\
57969&68.46621&5e-03&2.1&0.1&55878&6.56&A1V&1\\
72979&10.49512&8e-04&6.0&0.1&--&7.70&A4Vs&1\\
81771&11.81314&4e-04&3.6&0.1&--&7.76&A4V&1\\
82484&4.53561&8e-03&7.2&0.3&--&8.09&A3III/IV&2\\
92762&16.39493&8e-04&4.1&0.2&--&7.80&A8V&1\\
110080&18.55808&6e-04&2.7&0.1&--&7.41&A5V&1\\
121191&21.59957&3e-03&7.7&0.2&--&8.16&A5IV/V&3\\
156408&10.64459&3e-04&10.4&0.3&33894&8.27&A7V&2\\
163482&5.92257&3e-03&3.0&0.1&274692&6.82&A0III/IV&2\\
168651&15.66042&3e-04&2.5&0.2&--&7.40&A9III&3\\
170461&12.59065&5e-04&3.7&0.2&250253&6.98&A9IV&2\\
171819&13.55119&2e-04&2.4&0.3&--&5.84&A7IV/V&3\\
172995&13.17514&2e-04&3.3&0.1&--&6.81&A9IV&3\\
177523&13.11519&2e-04&2.6&0.2&--&7.49&A9/F0IV&3\\
177665&11.34302&2e-04&5.1&0.2&--&8.37&F2IV&3\\
189951&12.19356&5e-04&2.1&0.3&63419&5.25&A9IV&3\\
191585&15.46432&4e-04&4.4&0.1&--&6.92&A2/3IV&2\\
200203&19.31605&7e-04&2.1&0.3&--&7.35&A4/A5II/III&1\\
204352&14.35455&3e-05&2.9&0.2&--&8.40&A9V&1\\
208094&18.46880&4e-04&3.0&0.3&--&8.21&A2IV&1\\
216743&16.88458&3e-04&43.3&0.2&--&7.25&A3V&3
\enddata
\tablerefs{
(1) \citet{Houk75}, 
(2) \citet{Houk82}, 
(3) \citet{Houk78}.}
\end{deluxetable*}

\subsection{Other Variables} \label{othervars}

In addition to the variables discussed in the previous sections, bRing detected evidence of periodic pulsations representing several different 'other' classes of variability including ellipsoidal variables (ELLs), rotation periods (ROT), $\beta$ Cepheids (BCEPs), among others. bRing was particularly sensitive to low-amplitude (typically $\simeq$ 10 mmag) slowly pulsating B stars (SPBs) and long period variables (LPVs). In Table \ref{otherp}, we list 80 stars previously classified as variables in the VSX catalog along with their VSX and bRing variability parameters, and classification in the VSX catalog (last column). We discuss some of these stars further in \S \ref{prevother} if the period of a non-LPV star detected with bRing was significantly different than a previously published period or if the periodogram revealed new additional periods of interest. In Table \ref{othern}, a list of new period detections or variable classifications are provided for 25 stars using this system based on their spectral and pulsation properties and light curve shapes, and these stars are discussed further in \S \ref{newother}.

\subsubsection{Previously Known Variables}\label{prevother}

In comparing the periods determined using bRing data to those listed in either VSX or \citet{Rimoldini12}, we find that only 20 of the 80 (25\%) had completely different periods. 
We discuss the ones that showed period differences in this section. 

bRing was able to provide more precise period measurements for 5 of the LPV, SR, and SRD stars. Five of these stars had completely different periods from the low precision periods reported in VSX or \citet{Rimoldini12}. 

{\it HD 177171 ($\rho$ Tel)}: For the young F5V HD 177171, we detect a strong periodicity of 1.55258 $d$. However, \citet{Rimoldini12} quote a period of 0.71187 day, and VSX reports a period of 4.73687 day from \citet{Koen02}. 
Both estimates are based on the sparse {\it Hipparcos} time series photometry ($\sim$70 data points), whereas the bRing dataset has $>$10$^2$ more points and dense coverage.  
We do not detect significant peaks at the periods reported by either \citet{Rimoldini12} or \citet{Koen02} (in VSX). 

{\it HD 60168 (PS Pup)}: With bRing we detected a slightly different period (2.07742 d) for the ELL variable PS Pup compared to that published in VSX (1.34220 d). The periodogram for PS Pup does not have a significant period near the VSX period of 1.34220 $d$ and the detected bRing period is near the 2 $d$ alias of the sidereal systematic. The star and its periodicity detected with bRing is reported in Table \ref{otherp} because the expected shape of an ELL variable is recovered at this period versus the sinusoid expected from a sidereal alias (hence we believe the periodicity to be real). 

{\it HD 172416 and HD 189631}: The $\gamma$ Dor-type stars HD 172416 and HD 189631 both had different periods in bRing than they have reported in VSX. For HD 172416, the reported VSX period is 0.99787 $d$, which is close to the primary sidereal systematic. If this was a real signal in bRing, it could not be recovered due to the proximity to this dominant systematic. The 0.59900 $d$ signal for HD 172416 was not recovered at all in the bRing periodogram and the detrending routine should not affect this period. The periodogram for this star revealed the primary period of 0.70578 $d$ as well as additional significant periods that could be useful for future analysis.

{\it V946 Cen, HD 116862, and V846 Ara:} There were several Be stars detected by bRing. bRing measured a different period for the stars V946 Cen, HD 116862, and V846 Ara compared to that reported in VSX; these three stars also happen to be $\gamma$ Cas (GCAS) variables, which are known to be irregularly variable on the order of a decade. The time between the study of \citet{Rimoldini12} and bRing ($\sim$ 5 -- 7 years) could be enough time for periods to drift, causing the differences in observed periods. bRing's long baseline could also be a factor, picking up underlying shifts. In particular, V946 Cen and HD 116862 both show two dominant periods while V846 Ara only shows a single dominant period at 0.43861 $d$.

The RS CVn (RS) variables hosted the most discrepancies between observed bRing periods and prior studies. bRing detected completely different periods for $\rho$ Tel, HD 201247, and HD 209234; bRing detected near half the original period for HD 56142.

bRing detected 26 slowly pulsating B-type (SPB) stars (including 4 new ones from \S \ref{newother} and the reclassified classical Cepheid HD 136633 from \S \ref{sec:ceph}). These stars are characterized by their spectral type, location near the main-sequence (as observed for this sample in Figure \ref{cmd}), and periods ranging from just short of a day to several days \citep{Miglio07,DeCat07}. They are also known to exhibit multiple oscillations in their light curves \citep{Miglio07} and have even co-exhibited BCEP pulsations in a few rare cases \citep{DeCat07}.

The periodograms for all of the SPB variables were individually inspected for multiple periods or shorter periods that may indicate BCEP pulsations similar to the stars from \citep{DeCat07}. In general, the previously observed SPB variables showed multiple independent periods in their periodograms and bRing detected different primary periods for HD 85871 and HD 159041. Further
analysis of the SPB stars detected with bRing is beyond the scope of this work.

{\it HD 140566}: The final star that showed a different period than the reported VSX period was HD 140566 (previously discussed in \S \ref{sec:ds}). bRing detects a period at 0.08783 $d$ whereas VSX reports a 193.7000 $d$ period and classified the star as a detached EB (ESD). The bRing light curve does not indicate evidence of an eclipsing system. The period is more indicative of a $\delta$ Scuti, and multiple modes appear in the periodogram. Its spectral type (A5IV) and CMD position (from Figure \ref{cmd}) agree with a $\delta$ Scuti classification. Therefore, this star should be relabelled as a candidate $\delta$ Scuti.

\subsubsection{New Variables}\label{newother}

In Table \ref{othern}, we list 25 stars that had no known previously published periodicity or classification. A suggested classification was based on the period of the activity, the shape of the light curve, and the spectral type of the star. 

The stars HD 103285, HD 73141, and HD 99757 were classified as SPB variables based on their periods and locations in Figure \ref{cmd}. The star HD 76566 (IY Vel) was also observed by \citet{Lefevre09}, which also detected the 2.10650 $d$ period. However, they left the SPB classification as uncertain; the SPB nature of this star is confirmed with the bRing data. Due to bRing's long baseline, it was possible to detect several long-period variables out to the Nyquist limit of the observations (typically around 266 $d$). The population of these variables is composed largely of K giants. For the stars in Table \ref{othern}, the survey was able to either provide newly measured periods or provide more precise values over VSX periods. 

{\it HD 143098}: For the G5V star HD 143098, a 8.56567 $d$ period was detected that is indicative of rotation. The star also shows enhanced chromospheric activity \citep[logR'HK = -4.46, -4.59][]{BoroSaikia18} and
strong Li I absorption \citep[EW(Li I $\lambda$6707 = 60 m\AA;][]{Torres06}. Combining this rotation period with the B-V color (0.686\,$\pm$\,0.015) reported in {\it Hipparcos} \citep{ESA97}, and the age-rotation calibration of \citet{Mamajek08}, we estimate a gyrochronological age of HD 143098 of 0.6 Gyr. This is consistent with the other two indicators - both the Li absorption strength and the chromospheric activity are also very consistent with an age similar to that of the $\sim$0.6-0.8 Gyr-old Hyades \citep{Soderblom90,Mamajek08}.

\startlongtable
\begin{deluxetable*}{c|c|c|c|c|c|c|c|c|c|c|c}\tablecaption{Other Previously Classified Variables Detected With bRing \label{otherp}}
\tablehead{\colhead{Name} &\colhead{HD} & \colhead{P} & \colhead{$\sigma_{P}$} & \colhead{A} & \colhead{$\sigma_{A}$} &\colhead{VSX ID} & \colhead{P$_{VSX}$} & \colhead{$V$} & \colhead{SpT} & \colhead{Ref} & \colhead{Type}\\
\colhead{...} &\colhead{...} & \colhead{d} & \colhead{d} & \colhead{mmag}& \colhead{mmag}& \colhead{...} & \colhead{d} & \colhead{mag} & \colhead{...}& \colhead{...}& \colhead{(VSX)}}
\startdata
--&13397&48.32756&4e-02&10&0.2&280885&49.20000&7.74&K0III&1&ROT\\
UX For&17084&0.95197&1e-03&21.3&0.5&14260&0.95600&8.04&G5/8V+(G)&2&RS\\
TV Pic&30861&0.85175&3e-05&30.1&0.4&26362&0.85199&7.44&A2V&1&ELL\\
TU Pic&33331&1.14708&2e-04&9.0&0.2&26361&1.14686&6.90&B5III&1&SPB\\
R Pic&30551&141.89257&26e00&523.7&5.4&26334&168.00000&7.59&K2/K3II:pe&1&SR\\
YZ Men&34802&19.36778&8e-02&24.1&0.3&18686&19.58000&7.76&K1IIIp&3&RS\\
AB Dor&36705&0.51443&2e-04&10.6&0.2&13645&0.513900&6.93&K1III(p)&3&TTS/ROT\\
lam Col&39764&1.28660&1e-03&1.4&0.3&9602&1.28701&4.87&B5V&2&ELL\\
SZ Pic&39917&4.94895&4e-03&0.1&0.4&26359&4.95000&7.89&G8V&2&ELL\\
TY Pic&42504&48.82267&3e-01&7.8&0.3&26365&50.20000&7.70&G8/K0III+F&3&RS\\
V Pic&43518&166.32289&2e00&16.8&0.2&26338&180.00000&7.41&K2III&3&SR\\
AE Men&46291&12.14125&2e-02&13.9&0.2&18692&12.03000&8.25&K2III+F/G&3&RS\\
TZ Pic&46697&13.64700&3e-03&13.7&0.2&26366&13.68000&7.64&K1III/IVp&3&RS\\
V448 Car&49877&55.99252&1e-01&49.8&0.5&6201&--&5.61&K5III&3&SRD\\
--&56142&10.57968&4e-01&6.1&0.3&55829&21.16000&7.57&F6/F7V&2&RS\\
PS Pup&60168&2.07742&2e-03&8.1&0.4&26919&1.34220&6.62&B8V&2&ELL\\
V372 Car&64722&0.11540&1e-05&4.6&0.1&6125&0.11600&5.68&B1.5IV&3&BCEP\\
V413 Pup&66235&1.59433&9e-04&11.1&0.3&27015&1.59406&7.68&B9IV&1&SPB\\
V415 Pup&66503&0.84905&1e-04&7.5&0.2&27017&0.84892&8.22&B5V&1&SPB\\
QR Pup&69342&3.55155&7e-04&21.5&0.4&26928&3.55180&8.06&B3II&1&ELL\\
HV Vel&73340&2.66807&8e-04&6.2&0.1&37638&2.66745&5.78&ApSi&1&roAp\\
omi Vel&74195&2.79716&1e-03&7.3&0.2&37807&2.79759&3.59&B3IV&1&SPB\\
--&74422&0.74500&0.001&42.4&0.5&400557&0.74500&8.12&A3IV&3&ACEP\\
V473 Car&76640&0.95421&1e-04&8.1&0.1&6226&0.95399&6.35&B5V&3&SPB\\
OW Vel&76875&66.50249&2e+00&51.7&0.6&37726&64.54000&7.66&K2/3III+A/F&1&SRD\\
OY Vel&77653&1.48775&4e-04&6.6&0.1&37728&1.48782&5.03&B9&1&ACV\\
PR Vel&78405&1.23898&4e-04&10.6&0.3&37732&1.23890&8.26&B5IV&1&SPB\\
PS Vel&79039&1.07481&1e-04&9.4&0.2&37733&1.07460&6.82&B4V&1&SPB\\
V480 Car&81654&40.80297&5e-01&18.9&0.5&6233&40.00369\tablenotemark{a}&7.87&B2/3V(e)&3&BE + GCAS\\
QZ Vel&85871&5.71790&7e-03&5.2&0.2&37750&1.03108&6.49&B1V&3&SPB\\
V335 Vel&85953&3.75717&3e-03&6.1&0.1&37751&3.75520&5.94&B2V&4&SPB\\
--&88825&1.45753&3e-05&10.5&0.2&--&--&6.09&B4Ve&3&BE(SPB)\\
V514 Car&92287&2.90561&6e-04&3.8&0.1&6267&2.90457&5.88&B3IV&3&ELL\\
V431 Car&97152&1.61853&1e-05&6.7&0.2&6184&1.61853&8.07&WC7+O7V&5&E/WR\\
KQ Mus&100359&1.23848&3e-04&13.6&0.2&19926&1.23834&6.88&B7IV&3&SPB\\
V810 Cen&101947&151.28362&7e-01&16.0&0.2&8107&130.00000&5.01&F9Ia&6&LPV\\
DE Cru&104631&3.68406&3e-03&10.2&0.2&10896&3.68800&6.77&B1II&3&SPB\\
DF Cru&104705&1.13486&2e-04&13.8&0.2&10897&1.13480&7.81&B0.5III&3&SPB\\
V1123 Cen&108015&58.59871&1e+00&42.5&0.6&44225&60.60000&7.97&F3/5Ib/II&1&SRD\\
V946 Cen&112999&1.13651&1e-04&12.2&0.2&8243&0.08883\tablenotemark{a}&7.38&B6III(n)&7&BE+GCAS\\
--&116862&0.80112&7e-04&3.7&0.1&58241&2.87078\tablenotemark{a}&6.26&B3IV&1&BE+GCAS\\
--&118258&49.69547&1e+00&4.8&0.2&287154&50.50000&8.01&G6V&3&RS\\
DF Cir&124672&0.36907&1e-05&13.7&0.2&136640&0.367772&7.55&F6V&8&ELL\\
V1001 Cen&125104&6.73825&1e-02&10.1&0.3&8298&6.73600&7.29&B4IV/V&3&DPV/ELL\\
HX Lup&125721&3.08809&7e-04&5.0&0.1&17817&3.08809&6.11&B1III&1&ELL\\
V761 Cen&125823&8.81327&2e-02&6.5&0.3&8058&8.81710&4.41&B2V&2&SXARI\\
eta Cen&127972&0.64250&6e-05&7.9&0.3&8347&0.64247&2.33&B1Vn+A&9&GCAS+LERI\\
LS TrA&137164&44.46069&6e-01&39.9&0.4&36931&45.00000&7.47&K1/K2IVp&3&RS\\
HV Lup&137518&2.82838&2e-03&38.9&0.7&17815&--&7.74&B1/2(I/IIIN)&9&BE\\
LZ TrA&138521&0.57015&6e-05&7.4&0.2&36938&0.57019&8.04&B9IV&3&SPB\\
--&140566\tablenotemark{b}&0.08783&1e-05&3.3&0.2&415978&193.70000&8.28&A5IV&1&ESD(DSCT)\\
--&142542&92.14955&2e+00&4.1&0.2&412695&324.00000&6.29&F3/F5V&2&M\\
V374 Nor&147894&2.77589&3e-04&6.7&0.2&20334&2.72950&7.24&B5III&1&ELL\\
V918 Sco&149404&9.81300&5e-03&4.8&0.1&33734&9.81300&5.48&O9Ia&10&ELL\\
--&149455&1.28303&2e-04&6.8&0.2&59146&1.28217&7.69&B7III/IV&3&SPB\\
--&151158&0.18178&1e-05&6.3&0.2&225575&0.18178&8.21&B2Ib/II&1&BCEP\\
OV Aps&151665&0.92038&1e-04&9.7&0.2&834&0.92044&8.07&A7III&3&ACV\\
V846 Ara&152478&0.43861&2e-05&8.0&0.2&3649&0.60646\tablenotemark{a}&6.30&B3Vnpe&9&BE+GCAS\\
V847 Ara&152511&0.94205&1e-04&6.5&0.1&3650&0.94213&6.53&B5III&3&SPB\\
V884 Sco&153919&3.41141&4e-04&7.3&0.2&33700&3.41161&6.53&O5F&10&ELL+HMXB\\
--&155190&1.66574&7e-04&7.2&0.1&59550&1.66571\tablenotemark{a}&7.12&B7III/IV&3&SPB\\
V824 Ara&155555&1.68379&5e-05&6.5&0.1&3627&1.68160&6.87&K1Vp&3&RS\\
--&156853&1.15959&3e-04&4.1&0.1&--&1.15945&7.60&AP SI&1&ACV\\
--&159041&3.80296&3e-03&5.9&0.2&--&0.09432\tablenotemark{a}&8.04&B9Ib/II&1&SPB\\
V1092 Sco&163254&0.83175&2e-04&11.2&0.2&33908&0.83167&6.74&B5Vn&1&SPB\\
V692 CrA&166596&1.65010&3e-02&4.1&0.2&10563&1.67000&5.46&B2.5III&1&SXARI\\
--&172416&0.82698&3e-04&6.6&0.1&62979&0.99787&6.61&F5V&1&GDOR\\
V364 Pav&175008&0.57389&5e-05&6.6&0.1&25084&0.57389&6.80&B9IV/V&3&SPB\\
rho Tel&177171&1.55272&6e-05&4.7&0.1&63113&4.73687&5.17&F7V&1&RS\\
--&189631&0.70578&8e-05&6.8&0.2&63407&0.59900&7.54&A9V&1&GDOR\\
--&193677&0.42930&3e-03&3.2&0.1&281860&0.85857&7.60&A2V&3&ROT\\
--&201247&6.32432&2e-02&2.7&0.1&63923&1.26590\tablenotemark{a}&6.84&G0&11&RS\\
--&209234&4.67895&1e-01&4.6&0.1&--&0.089454\tablenotemark{a}&7.87&G3V&8&RS\\
CX Gru&214291&0.87089&1e-04&11.7&0.2&14793&0.87125&6.57&F7V&1&ELL\\
ksi Oct&215573&1.76908&3e-04&5.7&0.1&20485&1.76866&5.32&B6IV&5&SPB\\
--&216668&1.78959&4e-04&6.2&0.2&305847&1.78997&7.89&A1V&3&VAR\\
BP Gru&217522&0.18301&2e-02&1.7&0.1&14762&--&7.52&Ap(Si)CR&12&roAp\\
CF Oct&196818&20.46033&3e-02&20.4&0.3&20449&19.97000&7.90&K0IIIp&3&BY\\
SX Phe&223065&0.05496&6e-03&33.4&0.8&26236&0.05496&7.31&A2Vvar&1&SXPHE\\
V715 CrA&168403&4.80085&3e-03&13.8&0.2&10586&4.79920&6.78&A0II/III(p)&2&ACV
\enddata
\tablerefs{
(1) \citet{Houk78}, 
(2) \citet{Houk82}, 
(3) \citet{Houk75}, 
(4) \citet{Cucchiaro77}, 
(5) \citet{Shara09}, 
(6) \citet{Keenan89}, 
(7) \citet{Garrison77}, 
(8) \citet{Torres06}, 
(9) \citet{Levenhagen06}, 
(10) \citet{Sota14}, 
(11) \citet{Gray06}, 
(12) \citet{Levato96}.}
\tablenotetext{a}{\citet{Rimoldini12}}
\tablenotetext{b}{Reclassified}
\tablecomments{ACEP: Anomalous Cepheid, ACV: $\alpha^2$ CVn, BCEP: $\beta$ Cephei, BE: Be star, BY: BY-Draconis type, DPV: Double Periodic, DSCT: $\delta$ Scuti, ED/ESD: Detached EB, ELL: Ellipsoidal Variable, GCAS: $\gamma$ Cass-type, GDOR: $\gamma$ Dor-type, HMXB: High-mass x-ray binary, LERI: $\lambda$ Eri-type, LPV: Long Period, M: Mira-type, roAp: Chemically peculiar, rapidly oscillating A star, RS: RS CV-type, SDOR: S Doradus-type, SPB: Slowly Pulsating B-type, SXARI: SX Arietis-type, SXPHE: SXPHE-type variable, TTS: T Tauri Star, WR: Wolf-Rayet.}
\end{deluxetable*}
\startlongtable
\begin{deluxetable*}{c|c|c|c|c|c|c|c|c|c|c|c}\tablecaption{Other New Variables Detected With bRing \label{othern}}
\tablehead{\colhead{Name} &\colhead{HD} & \colhead{P} & \colhead{$\sigma_{P}$} & \colhead{A} & \colhead{$\sigma_{A}$} &\colhead{VSX ID} & \colhead{P$_{VSX}$} & \colhead{$V$} & \colhead{SpT} & \colhead{Ref} & \colhead{Type}\\
\colhead{...} &\colhead{...} & \colhead{d} & \colhead{d} & \colhead{mmag}& \colhead{mmag}& \colhead{...} & \colhead{d} & \colhead{mag} & \colhead{...}& \colhead{...}& \colhead{(VSX)}}
\startdata
--&4737&82.79415&4e00&5.1&0.1&--&--&6.27&G8III&1&LPV\\
--&5135&51.35633&3e-01&6.5&0.2&--&--&7.87&G3IV/V&1&LPV\\
--&6269&99.68567&9e00&13.4&0.8&--&--&6.28&G8IIICN...&2&LPV\\
--&11597&72.28817&4e00&20.7&1.1&--&--&8.15&F5V&2&LPV\\
--&32453&29.02423&4e-01&3.9&0.1&--&--&6.01&G8III&2&LPV\\
--&39937&47.04209&0.39153&10.4&0.2&41338&--&5.94&F7IV&3&LPV\\
--&53143&9.59515&6e-02&8.5&0.1&--&--&6.81&K0IV-V&3&ROT?\\
f Pup&61330&85.64948&1e-02&5.2&0.2&--&--&4.53&B8IV/V&2&LPV\\
--&73141&0.89316&2e-04&12.5&0.3&--&--&8.40&B7 III/IV&1&SPB\\
--&76006&17.97482&1e-01&4.0&0.1&--&--&7.32&F5/6 III&1&LPV\\
IY Vel&76566&2.10650&2e-03&3.6&0.1&37658&--&6.25&B3IV&1&SPB\\
--&87896&58.02840&1e00&6.2&0.1&--&--&6.91&G8III&3&LPV\\
--&90885&7.83318&3e-01&3.8&0.1&--&--&8.33&K2/K3III&4&ROT?\\
--&99757&1.13729&4e-04&12.5&0.3&--&--&8.16&B7 II/III&1&SPB\\
--&103285&1.33863&1e-04&6.6&0.3&411690&164.30000&8.23&B9.5V&3&SPB\\
--&129118&72.74121&14e00&5.4&0.1&--&--&6.79&K0III&1&LPV\\
KL Lup&135411&65.04632&26e00&15.5&0.4&17838&--&8.24&K5III&2&LPV\\
--&143098&8.56567&3e-04&3.5&0.2&--&--&7.64&G5V&4&ROT\\
--&144951&114.70571&2e-01&21.0&0.3&412789&138.00000&8.06&B3 V&3&LPV\\
--&149238&139.28268&6e00&32.5&0.3&36806&299.00000&8.04&F8V&3&LPV\\
--&156768&121.17629&6e00&4.7&0.1&--&--&5.86&G8Ib/II&3&LPV\\
--&167714&147.74923&24e00&5.0&0.1&--&--&5.94&K2III&3&LPV\\
--&179522&110.02770&6e00&10.0&0.1&63153&--&7.42&G8III/IV&1&LPV\\
--&192594&32.91187&6e-02&35.0&0.6&--&--&7.35&K3 III&2&LPV\\
--&198752&73.25503&1e00&11.0&0.2&63815&--&7.13&K4III&2&LPV
\enddata
\tablerefs{(1) \citet{Houk78}, (2) \citet{Houk82}, (3) \citet{Houk75}, (4) \citet{Torres06}.}
\tablecomments{LPV: Long Period, ROT: Rotation Period, SPB: Slowly Pulsating B star.}
\end{deluxetable*}

\subsection{Irregular Variables}

We also detected 17 irregular variables with the bRing photometry, which are listed in Table \ref{irr}. The table includes previously known irregular variables (Be stars, Mira variables) as well as stars whose bRing light curves showed evidence of variability, but for which no significant period could be converged on.

The majority of these irregular variables exhibited very short bright events (HD 3359, HD 68809, HD 69256, HD 72838, HD 91869, HD 127755, HD 128679, HD 205834). These events were only a few hours in nature and could be degenerate with isolated stochastic events in the bRing data. 

The light curve for HD 92063 shows three consecutive bumps in the brightness that extend as bright as 0.3 mag. 
The source of the bumps is likely due to the effects of the nova ASASSN-18fv (Nova Carina 2018, V906 Car)
\citep{Stanek18}, situated only 128\arcsec\, away from HD 92063. 
ASASSN-18fv was alternatively classified as either a classical nova \citep{Luckas18, Rabus18} or 
a young stellar object that underwent a burst of accretion \citep{Strader18}.
Given the proximity of the nova compared to the bRing pixel size, the light curve of HD 92063 is likely corrupted by ASASSN-18fv.

bRing also observed a few isolated dimming events in the stars HD 4229, HD 12440, HD 35324, and HD 222060; however, the events could not be verified as real, significant events due to their short duration and noisy characteristics. We also identified HD 36597, HD 51801, HD 139534, and HD 159468 as semi-regular variables in the bRing data. 

{\it HD 155806 (V1075 Sco) and HD 158864 (V830 Ara)}: V1075 Sco (O8V star classified as a Be in VSX) shows significant variability (amplitude of 0.05 mag) over the first 100 days of bRing data. After disappearing from the bRing data for about 100 more days, the star reappears as roughly constant. This is likely just a symptom of the irregularity exhibited by these stars. Since HD 155806 is not listed as a GCAS in any source, bRing simply classified this as an irregular star. V830 Ara showed several significant brightening events ($\simeq$ 0.2 mag) throughout the 500 days of observation. A representative period could not be converged on, so this was classified as irregular.

\startlongtable
\begin{deluxetable}{c|c|c|c|c|c}\tablecaption{Irregular Variables Detected With bRing \label{irr}}
\tablehead{\colhead{HD} &\colhead{VSX ID} & \colhead{P$_{VSX}$} & \colhead{$V$} & \colhead{SpT} & \colhead{Ref}\\
\colhead{...} & \colhead{...} & \colhead{d} & \colhead{mag} & \colhead{...}& \colhead{...}}
\startdata
3359&--&--&8.39&K0V&1\\
4229&--&--&6.8&K5III&2\\
12440&--&--&8.19&K2III&2\\
36597&--&--&3.86&K1II/III&3\\
51801&55670&224.71910&7.15&K2/K3III&2\\
68809&--&--&7.93&K0III&1\\
69256&--&--&8.17&K0III&1\\
72838&--&--&7.25&K1Ib:&3\\
91869&--&--&6.9&G8/K0III+..&2\\
92063&43533&--&5.08&K1III&2\\
127755&58566&--&7.66&K3 III&2\\
128679&--&--&7.76&K2 III&2\\
139534&36907&295.39108&7.8&K0II/III&2\\
155806&33891&3.20848\tablenotemark{a}&5.61&O8Ve&4\\
156468&33894&--&7.87&B2V:ne&3\\
158864&3633&--&8.17&B2 IB/IIeP&1\\
205834&--&--&8.1&K0 III&2\\
222060&--&--&5.99&K0II/III&2
\enddata
\tablenotetext{a}{\citet{Rimoldini12}}
\tablerefs{(1) \citet{Houk78}, (2) \citet{Houk75}, (3) \citet{Houk82}, (4) \citet{Sota14}.}
\end{deluxetable}

In \cite{Mamajek12}, the authors detected an unusually deep eclipsing event that took place over $\sim$50 $d$ in the J1407 system; this light curve has been subsequently modelled as a circumplanetary ring system \citep{vanWerkhoven14,Kenworthy15}. Follow-up observations and a study of the archival photometry have attempted to deduce the periodic nature of this event, but an additional event has yet to be confirmed \citep{Mentel18}. While searching for variables in the bRing data, evidence of such eclipses in other stars were also searched for. Unfortunately, no other eclipses of similar length or depth were detected.

\section{Conclusion} 

The bRing survey of $\beta$ Pictoris included nearly continuous photometric monitoring of 10,000+ bright variables in the southern sky
\citep{Stuik17, Mellon19AAS, Mellon19}. This paper reports on the variability of the bright
($V$ $\simeq$ 4-8 mag) stars observed during the bRing survey, and provides improved periods (and sometimes classifications) for many known variables, identifies new variables missed by previous surveys, and provides classifications for some VSX candidate variables. The light curves were also examined for any evidence of transits by circumstellar or circumplanetary dust disks or ring systems analogous to J1407 (V1400 Cen), but no such cases were observed.

Of the \Nstarstot analyzed in this survey, \Nstars stars were detected as variable (80\% were previously known and 20\% were new detections or classifications). These stars were separated by variability into several tables where identifying information and the bRing periods and amplitudes were provided. We provided a brief discussion on the stars whose periods or classifications deviated from those established by the VSX catalog or other surveys. We also provided brief discussion on the newly detected variables and the stars that should be reclassified to better fit their period, light curve shape, and spectral type. bRing was able to measure the O'Connell effect in 18 of the contact binaries in this survey; only 2 of these had previously shown the asymmetry in their light curves. This survey also detected 18 irregular variables, which were briefly discussed. The results from this  survey of the bRing time series photometry provides initial assessments of the variability parameters for bright southern stars, and may provide opportunities for further study to constrain the nature of these new and reclassified variable stars. 
\acknowledgements
SNM is a U.S. Department of Defense SMART scholar sponsored by the U.S. Navy through NIWC-Atlantic.
The results reported herein benefitted from collaborations and/or
information exchange within NASA's Nexus for Exoplanet System Science
(NExSS) research coordination network sponsored by NASA's Science
Mission Directorate.
Part of this research was carried out at the Jet Propulsion
Laboratory, California Institute of Technology, under a contract with
NASA.
The authors would like to acknowledge the support staff at both the South African 
Astronomical Observatory and Siding Spring Observatory for keeping both bRing stations maintained and running.
Construction of the bRing observatory to be sited at Siding Springs, Australia would
not be possible without a University of Rochester University Research Award,
help from Mike Culver and Rich Sarkis (UR), and generous donations of time,
services, and materials from Joe and Debbie Bonvissuto of Freight Expediters,
Michael Akkaoui and his team at Tanury Industries, Robert Harris and Michael
Fay at BCI, Koch Division, Mark Paup, Dave Mellon, and Ray Miller and the
Zippo Tool Room.
This research has made use of the International Variable Star Index (VSX) database, 
operated at AAVSO, Cambridge, Massachusetts, USA.
This research has made use of the VizieR catalogue access tool, CDS, Strasbourg, France (DOI: 10.26093/cds/vizier).
This research has made use of the SIMBAD database,
operated at CDS, Strasbourg, France.
We acknowledge with thanks the variable star observations from the AAVSO International Database contributed by observers worldwide and used in this research.
\facilities{bRing-SA, bRing-AU, AAVSO}
\software{Python 3.7.3 \citep{Python}, 
		  scipy \citep{Scipy}, 
          matplotlib \citep{Matplotlib},
          numpy \citep{Numpy},
          astropy \citep{astropy}
          }
\appendix
\startlongtable
\begin{deluxetable*}{c|c|c|c|c|c|c|c|c|c|c}
\tablecaption{Adopted and Calculated Stellar Parameters \label{cmdtab}}
\tablehead{\colhead{Name} &\colhead{$l$} & \colhead{$b$} & \colhead{$\varpi$} & \colhead{$B$} & \colhead{$V$} &\colhead{$(B-V)$} & \colhead{$E(B-V)$} & \colhead{$(B-V)_0$} & \colhead{$A_V$} & \colhead{$M_V$}\\
\colhead{...} &\colhead{$^\circ$} & \colhead{$^\circ$} & \colhead{mas} & \colhead{mag}& \colhead{mag}& \colhead{mag} & \colhead{mag} & \colhead{mag} & \colhead{mag}& \colhead{mag}}
\startdata
HD 3112&305.008486&-45.789165&7.162&6.359&6.109&0.250&0.005&0.245&0.015&0.369\\
HD 3359&309.594473&-67.800879&23.039&9.150&8.370&0.780&0.001&0.779&0.003&5.179\\
HD 3463&304.865029&-47.640183&5.377&8.240&8.040&0.200&0.008&0.192&0.025&1.668\\
HD 4229&303.190503&-31.421460&7.005&8.099&6.808&1.291&0.008&1.283&0.026&1.009\\
HD 4737&304.207519&-70.424430&7.996&7.175&6.277&0.898&0.002&0.896&0.006&0.785\\
HD 5135&302.466765&-65.577096&5.346&8.570&7.880&0.690&0.004&0.686&0.013&1.508\\
HD 6269&256.206003&-86.456746&7.841&7.213&6.283&0.930&0.002&0.928&0.006&0.749\\
HD 6882&297.833132&-61.714404&10.920&3.908&4.014&-0.106&0.002&-0.108&0.006&-0.801\\
HD 74422&278.644521&-12.886437&3.120&8.390&8.120&0.270&0.087&0.183&0.268&0.323\\
HD 8351&271.660105&-78.144661&7.234&6.960&6.698&0.262&0.002&0.260&0.006&0.989\\
HD 11413&280.699247&-64.277086&12.726&6.080&5.929&0.151&0.001&0.150&0.003&1.449\\
HD 11597&239.679217&-75.489274&12.648&8.610&8.150&0.460&0.001&0.459&0.003&3.657\\
HD 12284&246.808665&-73.257393&4.900&9.448&9.108&0.340&0.004&0.336&0.012&2.547\\
HD 12440&296.850833&-42.225420&2.537&9.500&8.200&1.300&0.030&1.270&0.098&0.124\\
HD 13397&272.939621&-63.724270&3.755&8.770&7.740&1.030&0.008&1.022&0.026&0.588\\
HD 16589&245.541212&-65.100855&17.948&6.995&6.483&0.512&0.001&0.511&0.003&2.750\\
HD 17084&244.720178&-64.175500&24.310&8.770&8.055&0.715&0.000&0.715&0.000&4.984\\
HD 17653&290.780175&-43.096853&17.029&7.120&6.660&0.460&0.002&0.458&0.006&2.810\\
HD 17755&283.036786&-49.418163&10.366&8.480&8.060&0.420&0.002&0.418&0.006&3.132\\
HD 20232&243.589022&-58.176492&12.584&7.040&6.880&0.160&0.001&0.159&0.003&2.376\\
HIP 21213&241.248424&-42.783693&7.500&7.800&7.630&0.170&0.001&0.169&0.003&2.002\\
HD 21765&276.560889&-47.102603&9.443&8.351&7.890&0.461&0.002&0.459&0.006&2.759\\
HD 25860&255.319180&-47.497443&7.153&6.826&6.615&0.211&0.002&0.209&0.006&0.881\\
HD 28837&245.182260&-43.294769&6.527&7.400&7.010&0.390&0.002&0.388&0.006&1.077\\
HD 30551&255.898732&-40.347787&0.870&8.990&6.350&2.640&0.012&2.628&0.042&-3.995\\
HD 30861&253.109163&-39.985308&4.691&7.580&7.440&0.140&0.003&0.137&0.009&0.787\\
HD 31407&264.090685&-38.718807&0.967&7.470&7.690&-0.220&0.013&-0.233&0.039&-2.422\\
HD 32453&243.737596&-37.381308&8.017&6.918&6.007&0.911&0.001&0.910&0.003&0.524\\
HD 32846&238.873189&-36.257079&10.170&6.607&6.314&0.293&0.001&0.292&0.003&1.348\\
HD 33331&250.226912&-36.738948&3.390&6.806&6.892&-0.086&0.005&-0.091&0.015&-0.472\\
HD 34349&275.285824&-34.809895&16.128&7.450&7.050&0.400&0.001&0.399&0.003&3.085\\
HD 34802&289.295356&-31.915956&5.409&8.850&7.770&1.080&0.030&1.050&0.097&1.339\\
HD 35324&282.662535&-33.019396&4.625&9.060&7.720&1.340&0.023&1.317&0.075&0.971\\
HIP 35815&251.367772&-11.077202&4.515&8.200&7.840&0.360&0.008&0.352&0.025&1.088\\
HD 36597&239.889723&-30.875387&12.461&5.010&3.870&1.140&0.001&1.139&0.003&-0.655\\
HD 36705&275.300830&-33.045550&65.320&7.856&6.999&0.857&0.000&0.857&0.000&6.074\\
HD 37350&271.733779&-32.774457&3.112&4.580&3.760&0.820&0.014&0.806&0.045&-3.819\\
HD 37513&287.845391&-31.095793&9.644&8.800&8.240&0.560&0.003&0.557&0.009&3.152\\
HD 37909&293.876606&-29.872461&6.892&8.510&8.260&0.250&0.008&0.242&0.025&2.427\\
HD 39244&252.963719&-29.692239&6.456&7.770&7.530&0.240&0.002&0.238&0.006&1.574\\
HD 39764&239.361225&-26.094313&9.666&4.720&4.870&-0.150&0.001&-0.151&0.003&-0.207\\
HD 39917&250.029969&-28.401647&5.307&8.660&7.906&0.754&0.003&0.751&0.010&1.521\\
HD 39937&265.481354&-30.410684&8.536&6.582&5.948&0.634&0.001&0.633&0.003&0.601\\
HD 41846&280.357024&-29.637540&5.192&8.470&8.110&0.360&0.014&0.346&0.044&1.643\\
HD 42504&262.654416&-27.941484&2.656&8.690&7.700&0.990&0.022&0.968&0.071&-0.250\\
HD 42933&263.302876&-27.683668&2.510&4.580&4.810&-0.230&0.027&-0.257&0.081&-3.273\\
HD 43518&268.921715&-27.999235&3.870&8.700&7.410&1.290&0.009&1.281&0.029&0.319\\
HD 43898&243.230633&-22.045071&6.925&8.160&7.870&0.290&0.002&0.288&0.006&2.066\\
HD 46291&282.691975&-27.681412&3.468&9.370&8.261&1.109&0.040&1.069&0.129&0.832\\
HD 46586&256.476067&-22.936854&8.882&8.320&8.030&0.290&0.001&0.289&0.003&2.770\\
HD 46697&268.279215&-25.562813&6.924&8.790&7.739&1.051&0.003&1.048&0.010&1.931\\
HD 46978&295.054679&-27.916876&5.090&8.480&8.160&0.320&0.073&0.247&0.226&1.468\\
HD 49877&265.125175&-22.645861&4.812&7.190&5.610&1.580&0.007&1.573&0.023&-1.001\\
HD 51801&281.264333&-25.361915&2.414&8.510&7.140&1.370&0.049&1.321&0.160&-1.107\\
HD 52993&246.244795&-13.439255&4.336&6.416&6.569&-0.153&0.007&-0.160&0.021&-0.267\\
HD 53143&271.654979&-22.592123&54.466&7.609&6.803&0.806&0.000&0.806&0.000&5.484\\
HD 54579&246.104700&-12.009698&15.469&8.940&8.029&0.911&0.001&0.910&0.003&3.973\\
HD 55173&242.507518&-9.724266&1.014&7.320&7.490&-0.170&0.049&-0.219&0.148&-2.628\\
HD 56142&248.137346&-11.516440&4.202&8.140&7.590&0.550&0.008&0.542&0.025&0.682\\
HD 56146&270.588989&-20.723588&2.261&8.050&8.090&-0.040&0.091&-0.131&0.276&-0.414\\
HD 56910&269.431682&-19.896621&6.452&7.100&6.840&0.260&0.005&0.255&0.015&0.873\\
HD 57969&267.448737&-18.489199&13.850&6.670&6.560&0.110&0.001&0.109&0.003&2.264\\
HD 58635&249.869298&-9.996722&5.520&7.100&6.810&0.290&0.006&0.284&0.019&0.501\\
HD 60168&249.247507&-8.119752&4.071&6.540&6.617&-0.077&0.010&-0.087&0.030&-0.365\\
HD 60559&253.130100&-9.785495&4.435&6.130&6.247&-0.117&0.011&-0.128&0.033&-0.552\\
HD 60649&265.536902&-15.781967&1.868&6.900&7.000&-0.100&0.109&-0.209&0.329&-1.972\\
HD 61330&248.978598&-6.669998&9.050&4.440&4.530&-0.090&0.002&-0.092&0.006&-0.693\\
HD 61644&273.717460&-18.682117&2.270&8.480&8.410&0.070&0.138&-0.068&0.419&-0.229\\
HD 63203&269.250591&-15.590387&1.914&8.300&8.320&-0.020&0.111&-0.131&0.336&-0.606\\
HD 63786&250.416015&-4.687935&6.539&5.886&5.936&-0.050&0.005&-0.055&0.015&-0.002\\
HD 64503&253.899409&-5.925270&5.030&4.301&4.474&-0.173&0.010&-0.183&0.030&-2.048\\
HD 64722&267.614913&-13.538668&2.313&5.529&5.680&-0.151&0.140&-0.291&0.420&-2.919\\
HD 65592&255.495365&-5.718947&0.802&8.130&7.370&0.760&0.356&0.404&1.110&-4.219\\
HD 65818&263.475590&-10.279223&3.400&4.240&4.410&-0.170&0.054&-0.224&0.163&-3.095\\
HD 66235&260.267567&-7.955540&2.508&7.540&7.670&-0.130&0.029&-0.159&0.088&-0.421\\
HD 66260&274.638128&-16.135648&5.388&7.760&7.460&0.300&0.016&0.284&0.050&1.068\\
HD 66503&258.189795&-6.404198&2.190&8.126&8.220&-0.094&0.025&-0.119&0.076&-0.154\\
HD 66623&251.745048&-2.180114&12.137&8.680&8.143&0.537&0.001&0.536&0.003&3.560\\
HD 66768&269.364185&-12.874937&2.557&6.660&6.690&-0.030&0.187&-0.217&0.564&-1.835\\
HD 68556&263.228698&-7.694949&0.616&8.780&8.160&0.620&0.120&0.500&0.376&-3.268\\
HD 68808&262.440856&-6.959550&1.217&6.350&5.760&0.590&0.066&0.524&0.207&-4.021\\
HD 68809&263.924453&-7.933014&2.491&9.070&7.930&1.140&0.042&1.098&0.136&-0.224\\
HD 68860&252.427475&-0.187155&0.584&8.000&6.700&1.300&0.202&1.098&0.653&-5.120\\
HD 69213&260.894205&-5.528365&9.858&6.980&6.700&0.280&0.002&0.278&0.006&1.663\\
HD 69256&262.640578&-6.653319&3.475&9.310&8.200&1.110&0.023&1.087&0.074&0.830\\
HD 69342&258.843122&-4.020185&--&8.240&8.060&0.180&--&--&--&--\\
HD 69879&249.199313&3.217249&7.055&7.475&6.422&1.053&0.004&1.049&0.013&0.652\\
HD 69882&259.498710&-3.908606&0.525&7.480&7.170&0.310&0.280&0.030&0.855&-5.085\\
HD 70999&256.821157&-0.612875&1.472&7.930&8.050&-0.120&0.113&-0.233&0.340&-1.450\\
HD 71302&260.502870&-2.888402&1.820&5.900&6.020&-0.120&0.065&-0.185&0.196&-2.876\\
HD 71487&257.618591&-0.538006&5.820&6.660&6.500&0.160&0.009&0.151&0.028&0.297\\
HD 71801&254.604570&2.030543&1.626&5.600&5.740&-0.140&0.062&-0.202&0.187&-3.391\\
HD 72275&275.249545&-12.285240&0.760&8.140&7.310&0.830&0.150&0.680&0.474&-3.760\\
HD 72698&274.678126&-11.515123&2.634&8.130&8.060&0.070&0.103&-0.033&0.314&-0.151\\
HD 72754&266.826694&-5.815158&0.582&9.090&8.880&0.210&0.231&-0.021&0.704&-2.999\\
HD 72838&266.517046&-5.480196&0.613&9.020&7.270&1.750&0.216&1.534&0.714&-4.505\\
HD 72878&271.659765&-9.204825&1.900&7.520&7.460&0.060&0.110&-0.050&0.335&-1.481\\
HD 72979&284.687144&-17.945936&4.154&7.900&7.700&0.200&0.045&0.155&0.138&0.654\\
HD 73141&267.079253&-5.553821&1.925&8.310&8.410&-0.100&0.098&-0.198&0.296&-0.464\\
HD 73340&268.266544&-6.176067&6.705&5.651&5.782&-0.131&0.005&-0.136&0.015&-0.101\\
HD 73502&262.882000&-1.910762&0.547&8.507&7.260&1.247&0.186&1.061&0.600&-4.651\\
HD 73699&259.795991&0.695547&1.004&7.620&7.600&0.020&0.087&-0.067&0.264&-2.656\\
HD 73882&260.181611&0.643141&2.170&7.590&7.190&0.400&0.036&0.364&0.112&-1.240\\
HD 74195&270.250715&-6.800198&6.610&3.440&3.630&-0.190&0.005&-0.195&0.015&-2.284\\
HD 74531&266.684501&-3.612202&1.331&7.080&7.230&-0.150&0.100&-0.250&0.301&-2.450\\
HD 74712&266.194768&-3.002906&0.258&9.580&8.320&1.260&0.230&1.030&0.741&-5.361\\
HD 74884&265.493657&-2.179077&0.409&9.190&8.330&0.860&0.312&0.548&0.980&-4.593\\
HD 75747&292.550759&-21.631893&10.096&6.280&6.070&0.210&0.005&0.205&0.015&1.075\\
HD 76006&268.697008&-3.333926&0.677&8.880&8.550&0.330&0.155&0.175&0.477&-2.773\\
HD 76566&265.641514&0.054159&2.601&6.100&6.260&-0.160&0.025&-0.185&0.075&-1.739\\
HD 76640&275.750650&-8.474583&4.580&6.257&6.353&-0.096&0.029&-0.125&0.088&-0.431\\
HD 76875&269.870923&-3.179071&2.681&8.830&7.720&1.110&0.031&1.079&0.100&-0.238\\
HD 77347&273.102827&-5.381860&4.675&7.840&7.580&0.260&0.019&0.241&0.059&0.870\\
HD 77464&271.159400&-3.523334&1.534&6.550&6.690&-0.140&0.054&-0.194&0.163&-2.544\\
HD 77581&263.058292&3.929854&0.384&7.370&6.870&0.500&0.142&0.358&0.442&-5.651\\
HD 77653&271.749887&-3.811274&8.850&5.171&5.295&-0.124&0.003&-0.127&0.009&0.021\\
HD 77669&265.662442&0.121400&1.955&8.050&8.100&-0.050&0.040&-0.090&0.121&-0.565\\
HD 78165&264.674744&3.317533&2.912&7.820&7.610&0.210&0.021&0.189&0.065&-0.134\\
HD 78405&272.634839&-3.739767&1.764&8.150&8.260&-0.110&0.049&-0.159&0.148&-0.655\\
HD 78763&285.437818&-14.797121&1.902&8.270&8.310&-0.040&0.066&-0.106&0.200&-0.494\\
HD 78801&271.863164&-2.558817&1.007&8.820&7.690&1.130&0.076&1.054&0.245&-2.539\\
HD 79039&269.132131&0.321652&2.273&6.692&6.812&-0.120&0.029&-0.149&0.088&-1.493\\
HD 81222&276.569793&-4.194508&0.911&8.290&7.570&0.720&0.217&0.503&0.680&-3.312\\
HD 81654&278.728196&-5.905462&0.916&7.880&7.880&0.000&0.176&-0.176&0.532&-2.843\\
HD 81771&288.159479&-15.200061&3.475&7.970&7.760&0.210&0.086&0.124&0.264&0.201\\
HD 82484&266.475505&8.320205&2.664&8.300&8.090&0.210&0.039&0.171&0.120&0.098\\
HD 82829&270.382574&4.770436&5.973&8.050&7.810&0.240&0.012&0.228&0.037&1.654\\
HD 84400&275.668170&1.411384&2.413&6.061&6.168&-0.107&0.043&-0.150&0.130&-2.050\\
HD 84416&285.886712&-10.535299&6.407&6.390&6.320&0.070&0.014&0.056&0.043&0.310\\
HD 84810&283.199003&-7.003808&0.777&4.330&3.400&0.930&0.156&0.774&0.496&-7.643\\
HD 85037&275.356375&2.830386&7.345&6.660&6.530&0.130&0.008&0.122&0.025&0.835\\
HD 85185&274.208898&4.519312&2.817&8.040&8.010&0.030&0.032&-0.002&0.098&0.161\\
HD 85871&279.410387&-0.870275&1.185&8.900&6.491&2.409&0.091&2.318&0.313&-3.453\\
HD 85953&276.869760&2.510135&1.684&5.777&5.932&-0.155&0.056&-0.211&0.169&-3.105\\
HD 86118&281.478387&-3.120892&1.341&6.470&6.640&-0.170&0.100&-0.270&0.301&-3.024\\
HD 86441&281.231910&-2.339378&1.033&7.490&7.520&-0.030&0.163&-0.193&0.492&-2.901\\
HD 87072&286.854973&-8.911832&0.858&9.070&8.290&0.780&0.155&0.625&0.489&-2.532\\
HD 87896&285.865065&-6.593891&6.670&7.820&6.910&0.910&0.015&0.895&0.048&0.983\\
HD 88278&295.294369&-18.743258&4.663&7.550&7.310&0.240&0.175&0.065&0.536&0.118\\
HD 88824&279.384303&4.265916&20.186&5.519&5.265&0.254&0.003&0.251&0.009&1.781\\
HD 88825&284.272213&-2.915781&1.455&5.997&6.087&-0.090&0.077&-0.167&0.233&-3.331\\
HD 89611&277.070166&9.457440&2.421&7.980&7.960&0.020&0.021&-0.001&0.064&-0.184\\
HD 89841&282.568898&1.474929&0.610&9.710&7.860&1.850&0.177&1.673&0.589&-3.803\\
HD 90000&279.472330&6.656732&1.299&7.420&7.560&-0.140&0.070&-0.210&0.211&-2.084\\
HD 90611&279.233514&8.442773&9.683&6.830&6.550&0.280&0.004&0.276&0.012&1.468\\
HD 90885&282.046388&4.407698&29.660&9.140&8.331&0.809&0.001&0.808&0.003&5.689\\
HD 90941&280.261842&7.510635&1.770&8.400&7.810&0.590&0.061&0.529&0.191&-1.142\\
HD 91519&273.992190&18.937293&4.197&7.980&7.700&0.280&0.030&0.250&0.093&0.722\\
HD 91869&285.492104&0.502151&2.070&7.790&6.910&0.880&0.046&0.834&0.147&-1.657\\
HD 92063&286.572675&-1.053904&13.237&6.249&5.088&1.161&0.010&1.151&0.032&0.664\\
HD 92287&285.630996&1.065114&2.666&5.749&5.876&-0.127&0.038&-0.165&0.115&-2.109\\
HD 92762&296.898420&-18.231703&7.730&8.060&7.810&0.250&0.009&0.241&0.028&2.223\\
HD 93130&287.568572&-0.859319&0.359&8.310&8.040&0.270&0.397&-0.127&1.202&-5.389\\
HD 93203&286.547631&1.212855&0.512&7.700&6.870&0.830&0.230&0.600&0.724&-5.308\\
HD 93486&298.409551&-20.324017&5.724&8.540&8.090&0.450&0.034&0.416&0.106&1.773\\
HD 93668&284.507820&6.151308&4.593&6.730&6.740&-0.010&0.020&-0.030&0.061&-0.010\\
HD 94924&292.748165&-8.238385&4.349&8.130&8.010&0.120&0.122&-0.002&0.372&0.830\\
HD 94985&285.137134&8.114793&9.124&6.061&5.898&0.163&0.006&0.157&0.018&0.680\\
HD 95109&289.057053&0.042811&0.479&7.210&6.110&1.100&0.400&0.700&1.267&-6.755\\
HD 95752&291.393564&-3.915900&0.314&7.421&6.970&0.451&0.447&0.004&1.364&-6.909\\
HD 95993&291.028404&-2.723612&2.260&8.620&8.180&0.440&0.245&0.195&0.755&-0.804\\
HD 97082&290.085281&1.472661&0.796&7.590&6.790&0.800&0.164&0.636&0.518&-4.224\\
HD 97152&290.946906&-0.488398&0.371&8.030&8.070&-0.040&0.352&-0.392&1.051&-5.133\\
HD 97485&291.472695&-1.111831&0.704&8.830&7.900&0.930&0.236&0.694&0.747&-3.610\\
HD 99757&289.751623&9.870446&2.160&8.060&8.160&-0.100&0.036&-0.136&0.109&-0.277\\
HD 100148&292.067042&4.264463&0.326&8.900&8.190&0.710&0.208&0.502&0.652&-4.894\\
HD 100213&294.807873&-4.144356&0.422&9.520&9.340&0.180&0.499&-0.319&1.495&-4.027\\
HD 100359&297.394601&-11.893218&1.867&7.081&6.884&0.197&0.323&-0.126&0.978&-2.739\\
HD 101947&295.179035&-0.644405&0.588&5.830&5.030&0.800&0.253&0.547&0.795&-6.917\\
HD 102541&290.030770&20.985574&8.604&8.180&7.940&0.240&0.006&0.234&0.019&2.595\\
HD 102682&291.846064&15.305595&3.918&8.789&8.240&0.549&0.052&0.497&0.163&1.042\\
HD 102893&295.568451&1.204261&1.120&8.280&8.250&0.030&0.147&-0.117&0.445&-1.949\\
HD 103285&296.026562&0.811795&1.347&8.310&8.240&0.070&0.138&-0.068&0.419&-1.533\\
HD 104036&300.068148&-15.248520&9.566&6.918&6.738&0.180&0.009&0.171&0.028&1.614\\
HD 104631&297.305591&0.165498&0.464&6.808&6.771&0.037&0.372&-0.335&1.114&-6.010\\
HD 104705&297.454691&-0.336299&0.432&9.560&9.110&0.450&0.386&0.064&1.181&-3.891\\
HD 105509&294.957397&17.886055&10.916&5.991&5.746&0.245&0.006&0.239&0.019&0.918\\
HD 106111&299.636182&-7.527775&1.131&7.010&6.170&0.840&0.227&0.613&0.716&-4.277\\
HIP 107231&349.866388&-48.057946&4.028&8.530&8.260&0.270&0.010&0.260&0.031&1.255\\
HD 107805&299.633990&1.064419&1.980&7.000&6.420&0.580&0.149&0.431&0.465&-2.562\\
HD 108015&298.254125&15.478414&0.212&8.379&7.960&0.419&0.064&0.355&0.199&-5.605\\
HD 108968&300.416306&3.351249&1.778&6.190&5.530&0.660&0.153&0.507&0.480&-3.700\\
HD 110080&302.015990&-7.694393&5.956&7.670&7.410&0.260&0.071&0.189&0.219&1.066\\
HD 110258&301.672146&3.053100&0.621&9.050&8.265&0.785&0.253&0.532&0.794&-3.564\\
HD 110311&302.104209&-6.550839&1.000&7.110&6.330&0.780&0.191&0.589&0.601&-4.271\\
HD 111984&303.343188&19.794944&6.910&7.510&7.280&0.230&0.028&0.202&0.086&1.391\\
HD 112044&303.316382&4.438949&1.022&7.340&6.580&0.760&0.218&0.542&0.685&-4.058\\
HD 112999&304.174646&2.175884&1.338&7.414&7.383&0.031&0.289&-0.258&0.869&-2.853\\
HD 114529&305.545614&2.845291&8.610&4.500&4.593&-0.093&0.019&-0.112&0.058&-0.790\\
HD 115823&307.408458&9.870003&8.166&5.314&5.443&-0.129&0.014&-0.143&0.042&-0.039\\
HD 116862&308.917726&13.072689&1.255&6.126&6.257&-0.131&0.075&-0.206&0.226&-3.476\\
HD 117399&307.686744&0.918454&0.484&7.190&6.490&0.700&0.451&0.249&1.394&-6.481\\
HD 118258&309.271499&6.154131&9.360&8.840&8.040&0.800&0.014&0.786&0.045&2.852\\
HD 118769&309.462365&4.637572&0.537&8.720&7.300&1.420&0.225&1.195&0.731&-4.780\\
HD 119888&309.107132&-1.669995&0.540&7.880&7.880&0.000&0.308&-0.308&0.924&-4.382\\
HD 120400&310.839533&4.377956&0.841&8.440&7.690&0.750&0.161&0.589&0.507&-3.192\\
HD 121191&312.441366&8.159707&7.570&8.400&8.160&0.240&0.020&0.220&0.062&2.494\\
HD 121291&314.795173&16.737341&0.996&8.640&7.900&0.740&0.054&0.686&0.171&-2.279\\
HD 122314&309.986969&-4.857728&6.034&7.970&7.620&0.350&0.063&0.287&0.195&1.328\\
HD 122844&313.625108&6.651686&7.752&6.433&6.204&0.229&0.026&0.203&0.080&0.571\\
HD 123720&314.119582&5.685342&6.262&7.980&7.750&0.230&0.044&0.186&0.136&1.598\\
HD 124195&314.733924&6.351017&3.726&6.130&6.090&0.040&0.185&-0.145&0.560&-1.614\\
HD 124672&310.895742&-6.521600&15.482&8.070&7.550&0.520&0.008&0.512&0.025&3.474\\
HD 124689&314.132939&3.145630&9.267&7.650&7.290&0.360&0.020&0.340&0.062&2.063\\
HD 125104&315.068011&4.908926&1.529&7.360&7.300&0.060&0.266&-0.206&0.802&-2.580\\
HD 125721&318.197805&11.829894&1.047&8.840&8.490&0.350&0.114&0.236&0.352&-1.762\\
HD 125823&321.565642&20.022614&7.130&4.240&4.420&-0.180&0.013&-0.193&0.039&-1.354\\
HD 126859&316.399573&4.142315&7.032&7.189&6.965&0.224&0.063&0.161&0.194&1.006\\
HD 127297&316.444594&3.307587&1.340&7.840&6.930&0.910&0.284&0.626&0.896&-3.331\\
HD 127755&315.349490&-0.303711&0.710&9.430&7.630&1.800&0.340&1.460&1.120&-4.233\\
HD 127972&322.773998&16.669138&10.670&2.120&2.310&-0.190&0.009&-0.199&0.027&-2.576\\
HD 128679&308.950206&-16.205993&2.853&9.090&7.760&1.330&0.077&1.253&0.251&-0.214\\
HD 129094&314.416788&-4.710796&9.186&9.920&9.460&0.460&0.000&0.460&0.000&4.276\\
HD 129118&323.658227&15.582768&6.354&7.780&6.790&0.990&0.014&0.976&0.045&0.760\\
HD 130233&313.973805&-7.214961&1.016&8.290&7.440&0.850&0.180&0.670&0.569&-3.095\\
HD 130701&315.825304&-4.013205&1.745&6.660&5.960&0.700&0.133&0.567&0.418&-3.249\\
HD 131638&325.032748&12.542091&2.462&8.330&8.320&0.010&0.062&-0.052&0.189&0.088\\
HD 132247&323.127179&7.800860&4.932&8.270&8.090&0.180&0.052&0.128&0.160&1.396\\
HD 133880&329.181429&15.214997&9.652&5.650&5.790&-0.140&0.011&-0.151&0.033&0.680\\
HD 135240&319.688221&-2.911210&1.556&5.030&5.090&-0.060&0.152&-0.212&0.458&-4.408\\
HD 135411&332.251211&16.998319&1.070&9.900&8.220&1.680&0.097&1.583&0.321&-1.954\\
HD 135592&316.976839&-7.757915&1.475&6.960&6.390&0.570&0.116&0.454&0.363&-3.128\\
HD 135876&330.846349&13.954524&6.974&5.495&5.604&-0.109&0.012&-0.121&0.036&-0.215\\
HD 136633&319.961709&-4.261537&1.143&8.280&8.230&0.050&0.171&-0.121&0.518&-1.998\\
HD 137164&319.624925&-5.332615&8.763&9.180&8.140&1.040&0.019&1.021&0.061&2.792\\
HD 137518&329.807083&9.395976&0.582&7.830&7.750&0.080&0.178&-0.098&0.540&-3.964\\
HD 137626&318.422350&-7.641804&0.889&8.610&7.800&0.810&0.141&0.669&0.446&-2.902\\
HD 138521&317.798582&-9.522212&3.742&8.050&8.040&0.010&0.062&-0.052&0.189&0.717\\
HD 139534&319.148827&-8.694991&4.067&8.810&7.810&1.000&0.065&0.935&0.208&0.648\\
HD 140566&332.392135&7.543717&3.791&8.510&8.290&0.220&0.064&0.156&0.197&0.987\\
HD 142049&323.988673&0.015545&19.510&6.210&5.850&0.360&0.005&0.355&0.016&2.286\\
HD 142542&342.704106&16.465330&18.766&6.718&6.284&0.434&0.003&0.431&0.009&2.642\\
HD 142941&322.129487&-8.223853&1.075&7.190&6.410&0.780&0.092&0.688&0.291&-3.724\\
HD 142994&338.396060&10.867654&6.347&7.460&7.170&0.290&0.011&0.279&0.034&1.149\\
HD 143028&321.769239&-8.741536&0.706&7.720&7.800&-0.080&0.105&-0.185&0.317&-3.273\\
HD 143098&340.463665&12.953532&30.331&8.330&7.640&0.690&0.002&0.688&0.006&5.043\\
HD 143232&338.403332&10.406861&6.784&6.890&6.660&0.230&0.008&0.222&0.025&0.793\\
HD 143999&323.231855&-8.042597&0.877&8.430&7.890&0.540&0.111&0.429&0.346&-2.741\\
HD 144951&328.159706&-3.508135&0.940&8.080&8.070&0.010&0.251&-0.241&0.755&-2.821\\
HD 146323&327.753558&-5.403478&1.062&7.490&6.490&1.000&0.192&0.808&0.611&-3.992\\
HD 147170&333.339110&-0.575511&4.619&8.895&8.263&0.632&0.101&0.531&0.317&1.269\\
HD 147683&344.856604&10.088824&3.387&7.160&7.050&0.110&0.355&-0.245&1.068&-1.369\\
HD 147894&335.606579&0.682539&6.090&7.241&7.312&-0.071&0.081&-0.152&0.245&0.990\\
HD 148891&315.735663&-18.647736&3.946&8.050&8.000&0.050&0.044&0.006&0.134&0.846\\
HD 149238&324.027972&-11.927103&18.618&8.580&8.040&0.540&0.006&0.534&0.019&4.371\\
HD 149404&340.537543&3.005780&0.760&5.880&5.520&0.360&0.442&-0.082&1.342&-6.419\\
HD 149450&338.788613&1.271181&1.121&8.229&8.239&-0.010&0.262&-0.272&0.787&-2.300\\
HD 149455&331.781126&-5.212324&4.271&7.710&7.690&0.020&0.149&-0.129&0.451&0.391\\
HD 149668&327.191739&-9.577443&2.115&7.710&7.610&0.100&0.079&0.021&0.241&-1.004\\
HD 149715&330.389958&-6.781829&3.569&9.420&8.330&1.090&0.135&0.955&0.433&0.659\\
HD 149779&339.877907&1.797286&1.154&7.740&7.560&0.180&0.306&-0.126&0.927&-3.056\\
HD 151158&341.876698&1.450784&0.856&8.480&8.250&0.230&0.419&-0.189&1.265&-3.353\\
HD 151475&338.875985&-1.588860&0.992&8.170&8.060&0.110&0.266&-0.156&0.804&-2.762\\
HD 151564&343.130554&1.910493&0.737&8.100&7.980&0.120&0.438&-0.318&1.313&-3.995\\
HD 151665&315.626975&-20.092090&2.907&8.320&8.070&0.250&0.056&0.194&0.173&0.215\\
HD 151890&346.115149&3.913990&3.726&2.820&2.980&-0.160&0.181&-0.341&0.542&-4.705\\
HD 152333&343.827246&1.374939&0.735&8.930&8.840&0.090&0.434&-0.344&1.299&-3.128\\
HD 152478&336.782978&-4.635986&3.222&6.310&6.330&-0.020&0.244&-0.264&0.733&-1.863\\
HD 152511&328.996568&-10.888160&4.900&6.468&6.530&-0.062&0.058&-0.120&0.176&-0.195\\
HD 152667&344.531027&1.457096&0.607&6.510&6.220&0.290&0.436&-0.146&1.319&-6.183\\
HD 152901&346.897945&3.025074&3.115&7.450&7.390&0.060&0.187&-0.127&0.566&-0.709\\
HD 153004&350.412633&5.666730&2.270&7.370&6.610&0.760&0.272&0.488&0.852&-2.462\\
HD 153140&340.596082&-2.424087&0.926&7.850&7.500&0.350&0.327&0.023&0.999&-3.665\\
HD 153747&347.139655&1.962184&5.469&7.540&7.420&0.120&0.033&0.087&0.101&1.008\\
HD 153919&347.754424&2.173487&0.549&6.780&6.510&0.270&0.402&-0.132&1.217&-6.011\\
HD 154339&340.795488&-3.813101&0.837&9.740&9.190&0.550&0.345&0.205&1.064&-2.259\\
HD 155190&328.390770&-13.744251&4.083&7.100&7.130&-0.030&0.060&-0.090&0.182&0.003\\
HD 155550&352.960668&3.540731&0.641&8.130&8.070&0.060&0.328&-0.268&0.986&-3.881\\
HD 155555&324.898586&-16.297357&32.780&7.499&6.723&0.776&0.003&0.773&0.010&4.292\\
HD 155775&348.796715&0.145558&0.877&8.660&8.610&0.050&0.306&-0.256&0.920&-2.595\\
HD 155781&330.235394&0.180648&3.986&7.550&7.420&0.130&0.059&0.071&0.181&0.242\\
HD 156408&349.316308&-0.389699&3.378&8.650&8.270&0.380&0.192&0.188&0.592&0.322\\
HD 156623&343.479370&-4.832280&8.948&7.350&7.260&0.090&0.008&0.082&0.025&1.994\\
HD 156768&333.053617&-12.071444&2.980&7.016&5.872&1.144&0.079&1.065&0.255&-2.012\\
HD 156853&340.014519&-7.486891&2.747&7.560&7.600&-0.040&0.116&-0.156&0.351&-0.557\\
HD 156979&343.510397&-5.220728&0.977&7.670&6.740&0.930&0.245&0.685&0.775&-4.086\\
HD 157321&327.178343&-16.158987&3.939&8.350&8.020&0.330&0.051&0.279&0.158&0.839\\
HD 158155&354.660584&0.814982&0.712&8.720&8.330&0.390&0.389&0.001&1.187&-3.594\\
HD 158186&355.906751&1.596463&0.940&7.070&7.040&0.030&0.268&-0.238&0.807&-3.902\\
HD 158443&354.359038&0.173160&0.920&8.780&7.910&0.870&0.298&0.572&0.937&-3.207\\
HD 159041&342.732780&-8.188957&1.634&7.990&8.040&-0.050&0.079&-0.129&0.239&-1.133\\
HD 159441&335.178910&-13.204213&9.550&7.690&7.360&0.330&0.008&0.322&0.025&2.235\\
HD 159654&349.029646&-4.865982&0.840&7.990&7.260&0.730&0.199&0.531&0.625&-3.744\\
HD 160589&356.568493&-1.268251&5.326&8.180&7.850&0.330&0.051&0.279&0.158&1.324\\
HD 161592&1.166282&0.209286&3.431&5.340&4.540&0.800&0.173&0.627&0.546&-3.328\\
HD 161783&338.940199&-13.195838&1.973&5.600&5.710&-0.110&0.122&-0.232&0.367&-3.182\\
HD 162102&356.489248&-3.419668&0.798&8.790&7.510&1.280&0.280&1.000&0.901&-3.882\\
HD 163181&358.125130&-3.774663&0.511&6.990&6.610&0.380&0.372&0.008&1.135&-5.984\\
HD 163254&349.894016&-8.614876&1.730&6.632&6.728&-0.096&0.083&-0.179&0.251&-2.333\\
HD 163482&355.219541&-5.828797&7.071&6.870&6.830&0.040&0.024&0.016&0.073&1.004\\
HD 163708&354.530336&-6.509347&5.712&7.155&7.089&0.066&0.033&0.033&0.101&0.772\\
HD 164975&1.575794&-3.979580&1.180&5.470&4.690&0.780&0.119&0.661&0.376&-5.327\\
HD 166596&351.887336&-10.965792&1.643&5.298&5.462&-0.164&0.053&-0.217&0.160&-3.619\\
HD 167231&357.282255&-8.844732&3.292&7.520&7.420&0.100&0.050&0.050&0.153&-0.145\\
HD 167714&313.739664&-25.774726&9.226&7.133&5.946&1.187&0.007&1.180&0.023&0.748\\
HD 168403&354.693942&-11.317882&4.582&6.900&6.790&0.110&0.027&0.083&0.083&0.012\\
HD 168651&346.130512&-15.563721&9.614&7.700&7.400&0.300&0.008&0.292&0.025&2.290\\
HD 168740&331.834147&-21.150130&14.147&6.313&6.122&0.191&0.006&0.185&0.018&1.857\\
HD 170461&357.594735&-12.142131&10.493&7.280&6.980&0.300&0.006&0.294&0.019&2.066\\
HD 171577&352.316173&-15.758740&3.472&7.780&7.750&0.030&0.037&-0.007&0.113&0.340\\
HD 171819&347.618520&-17.856894&10.163&6.058&5.840&0.218&0.008&0.210&0.025&0.850\\
HD 172416&347.960987&-18.296431&12.444&7.080&6.620&0.460&0.007&0.453&0.022&2.073\\
HD 172995&347.659912&-18.979164&6.290&7.010&6.810&0.200&0.018&0.182&0.055&0.748\\
HD 173344&332.368984&-23.771926&6.866&7.590&7.410&0.180&0.014&0.166&0.043&1.551\\
HD 173794&344.012056&-20.878022&6.220&7.350&7.120&0.230&0.020&0.210&0.062&1.027\\
HD 174139&338.884357&-22.668354&3.165&8.240&8.170&0.070&0.047&0.023&0.144&0.528\\
HD 174632&5.157964&-13.722970&4.227&6.607&6.638&-0.031&0.056&-0.087&0.170&-0.402\\
HD 174694&328.286503&-25.387694&5.199&5.080&4.400&0.680&0.016&0.664&0.051&-2.071\\
HD 175008&323.165007&-26.345258&5.397&6.750&6.790&-0.040&0.011&-0.051&0.033&0.417\\
HD 177171&344.540829&-23.346929&16.990&5.700&5.174&0.526&0.006&0.520&0.019&1.306\\
HD 177523&349.613583&-22.216656&7.289&7.780&7.490&0.290&0.012&0.278&0.037&1.766\\
HD 177665&351.911604&-21.656600&5.185&8.690&8.360&0.330&0.025&0.305&0.078&1.856\\
HD 177776&340.946298&-24.572862&3.143&8.150&8.120&0.030&0.038&-0.008&0.116&0.491\\
HD 179522&349.644068&-23.582311&8.066&8.330&7.430&0.900&0.010&0.890&0.032&1.931\\
HD 184035&358.915878&-24.806037&6.696&6.000&5.910&0.090&0.019&0.071&0.058&-0.019\\
HD 185139&353.528055&-27.057738&9.902&6.530&6.260&0.270&0.007&0.263&0.022&1.217\\
HD 187418&352.217174&-29.447853&3.583&8.590&8.310&0.280&0.019&0.261&0.059&1.023\\
HD 189631&358.788751&-30.384035&11.087&7.840&7.540&0.300&0.006&0.294&0.019&2.746\\
HD 189951&354.800696&-31.301714&5.250&8.100&7.830&0.270&0.012&0.258&0.037&1.394\\
HD 191585&331.765945&-33.205593&5.120&7.060&6.920&0.140&0.014&0.126&0.043&0.423\\
HD 192316&314.682531&-30.934019&7.588&7.770&7.550&0.220&0.007&0.213&0.022&1.929\\
HD 192594&2.167096&-32.659315&3.073&8.700&7.340&1.360&0.025&1.335&0.082&-0.304\\
HD 193174&9.694296&-31.884618&7.106&7.500&7.250&0.250&0.013&0.237&0.040&1.468\\
HD 193677&343.593132&-35.073332&8.396&7.760&7.600&0.160&0.008&0.152&0.025&2.196\\
HD 198592&353.432167&-39.866271&6.508&7.780&7.580&0.200&0.008&0.192&0.025&1.623\\
HD 198736&329.508381&-37.301504&5.965&8.630&8.340&0.290&0.008&0.282&0.025&2.193\\
HD 198752&6.960965&-39.376813&3.287&8.620&7.140&1.480&0.018&1.462&0.059&-0.335\\
HD 200203&337.197874&-39.987120&5.505&7.530&7.350&0.180&0.009&0.171&0.028&1.026\\
HD 200475&319.771576&-35.486409&4.474&8.080&7.820&0.260&0.021&0.239&0.065&1.008\\
HD 200670&7.258495&-41.851515&12.481&8.340&7.810&0.530&0.004&0.526&0.013&3.279\\
HD 201247&342.588946&-41.727507&28.760&7.700&7.100&0.600&0.004&0.596&0.013&4.381\\
HD 201292&310.233667&-31.637037&4.720&8.510&8.190&0.320&0.094&0.226&0.290&1.270\\
HD 201427&345.509006&-42.280150&20.650&7.730&7.068&0.662&0.005&0.657&0.016&3.627\\
HD 203244&324.896406&-38.909688&48.062&7.700&6.970&0.730&0.001&0.729&0.003&5.376\\
HD 204352&339.381016&-43.994561&5.251&8.660&8.400&0.260&0.008&0.252&0.025&1.977\\
HD 204370&347.479373&-45.467633&4.886&7.820&7.520&0.300&0.009&0.291&0.028&0.937\\
HD 205834&307.420813&-30.917878&4.261&9.340&8.110&1.230&0.093&1.137&0.301&0.956\\
HD 205877&344.098347&-46.530732&5.135&6.784&6.202&0.582&0.008&0.574&0.025&-0.271\\
HD 208094&305.516566&-29.818144&4.135&8.440&8.210&0.230&0.081&0.149&0.249&1.043\\
HD 208614&352.610492&-51.199348&3.686&7.870&7.720&0.150&0.009&0.141&0.028&0.525\\
HD 209234&331.450421&-46.325930&23.352&8.480&7.870&0.610&0.003&0.607&0.009&4.702\\
HD 209295&326.230432&-44.109185&8.528&7.570&7.320&0.250&0.005&0.245&0.015&1.959\\
HD 210572&338.681592&-50.347903&13.575&8.238&7.714&0.524&0.004&0.520&0.013&3.365\\
HD 212661&332.056482&-50.187132&6.910&7.080&6.910&0.170&0.005&0.165&0.015&1.092\\
HD 213669&336.103568&-53.075908&8.779&7.620&7.420&0.200&0.004&0.196&0.012&2.125\\
HD 214291&359.582222&-59.601060&10.317&7.120&6.590&0.530&0.003&0.527&0.009&1.648\\
HD 215573&309.029149&-35.525946&6.326&5.180&5.311&-0.131&0.009&-0.140&0.027&-0.711\\
HD 216668&317.189250&-45.028130&3.871&7.980&7.880&0.100&0.015&0.085&0.046&0.773\\
HD 216743&352.070787&-61.942147&9.768&7.413&7.247&0.166&0.003&0.163&0.009&2.187\\
HD 217522&346.722689&-61.851320&9.715&7.990&7.520&0.470&0.003&0.467&0.009&2.448\\
HD 218090&325.729448&-53.525332&5.474&8.420&8.130&0.290&0.006&0.284&0.019&1.803\\
HD 219301&326.385409&-55.844349&10.907&6.840&6.560&0.280&0.003&0.277&0.009&1.739\\
HD 220633&345.169775&-66.525355&2.200&8.770&8.290&0.480&0.009&0.471&0.028&-0.026\\
HD 222060&308.224592&-39.498589&6.123&6.900&5.983&0.917&0.009&0.908&0.029&-0.111\\
HD 223065&341.382354&-70.363573&12.000&7.400&7.120&0.280&0.002&0.278&0.006&2.510
\enddata
\end{deluxetable*}

\bibliography{mellon.bbl}{}
\bibliographystyle{aasjournal}

\end{document}